\documentclass[format=acmsmall]{acmart}

\usepackage[utf8]{inputenc}

\usepackage{subcaption}

\usepackage{booktabs}
\usepackage{marvosym}
\usepackage{color}
\usepackage{graphicx}
\usepackage{hyperref}
\usepackage{microtype}
\usepackage{forest}

\usepackage[T1]{fontenc}

\usepackage{tikz}
\usetikzlibrary{calc}
\usetikzlibrary{fit}
\usetikzlibrary{calc}
\usetikzlibrary{shapes}
\usetikzlibrary{shapes.geometric}
\usetikzlibrary{shapes.multipart}
\usetikzlibrary{shapes.gates.logic.US}
\usetikzlibrary{arrows}
\usetikzlibrary{arrows.meta}

\tikzset{
  big edge/.style={
    green,
    thick,
  },
  big edgep/.style={
    big edge,
    -{Circle[fill=black,black,width=2,length=2,sep=-1]}
  },
  big pedge/.style={
    big edge,
    {Circle[fill=black,black,width=2,length=2,sep=-1]}-
  },
  big pedgep/.style={
    big edge,
    {Circle[fill=black,black,width=2,length=2,sep=-1]}-{Circle[fill=black,black,width=2,length=2,sep=-1]}
  },
  big edgec/.style={
    big edge,
    -{Bar[fill=green,green,width=4,length=0,sep=0]}
  },
  big pedgec/.style={
    big edge,
    {Circle[fill=black,black,width=2,length=2,sep=-1]}-{Bar[fill=black,black,width=4,length=0,sep=0]}
  },
  big region/.style={
    draw,
    rectangle,
    rounded corners=1.5,
    dashed,
    dash pattern=on 1pt off 1pt,
    thin,
    gray,
  },
  big site/.style={
    big region,
    fill=gray!60,
    text=black,
  },
  big react/.style={
    black,
    thick,
    -stealth,
    line width=3,
    shorten <=3,
    shorten >=3,
  },
  big react rev/.style={
    black,
    thick,
    stealth-stealth,
    line width=3,
    shorten <=3,
    shorten >=3,
  },
  big inst map/.style={
    thick,
    -stealth,
    blue,
    dashed
  },
  lbl/.style={
    font=\tiny\sf,
    inner sep=1,
  },
  lbl conc/.style={
    font=\tiny,
    inner sep=1,
  }
}
 \tikzset {
  room/.style={
    draw,
    label={[inner sep=0.5]north:{\sf\tiny Room}},
  },
  floor/.style={
    draw,
    label={[inner sep=0.5]north:{\sf\tiny Floor}},
  },
  building/.style={
    draw,
    label={[inner sep=0.5]north:{\sf\tiny Building}},
  },
  camera/.style={
    draw,
    ellipse,
    label={[inner sep=0.5]north:{\sf\tiny Camera}},
  },
  ctrlPanel/.style={
    draw,
    rounded corners=2,
    label={[inner sep=0.5]north:{\sf\tiny CtrlPanel}},
  },
  server/.style={
    draw,
    rounded corners=2,
  },
  composeColour/.style={
    orange
  },
  composeLink/.style={
    -latex,
    composeColour,
    thick,
    dashed
  }
}

\newcommand{\rr}[1]{\texttt{#1}}
\newcommand{\rrP}[2]{{\texttt{#1(}#2\texttt{)}}}

\newcommand\one{\biginline{1}}

\DeclareMathOperator{\react}{\mathrel{\frac{\raisebox{0.75mm}{\begin{scriptsize}\ensuremath{\hspace*{1mm}\ \hspace*{1mm}}\end{scriptsize}}}{}} \joinrel{\!\!\vartriangleright}}

\DeclareMathOperator{\rrul}{\mathrel{\frac{\raisebox{0.75mm}{\begin{scriptsize}\ensuremath{\hspace*{1mm}\ \hspace*{1mm}}\end{scriptsize}}}{}} \joinrel{\!\!\blacktriangleright}}

\newcommand{\rrulp}[1]{\operatorname{\mathrel{\frac{\raisebox{0.75mm}{\begin{scriptsize}\ensuremath{\hspace*{1mm}\ #1 \hspace*{1mm}}\end{scriptsize}}}{}} \joinrel{\!\!\blacktriangleright}}}
\newcommand{\rrula}[2]{\operatorname{\mathrel{\frac{\raisebox{0.75mm}{\begin{scriptsize}\ensuremath{\hspace*{1mm}\ #1 \hspace*{1mm}}\end{scriptsize}}}{\begin{scriptsize}\ensuremath{\hspace*{1mm}\ #2 \hspace*{1mm}}\end{scriptsize}}}\joinrel{\!\!\blacktriangleright}}}

\newcommand{\tipbox}[1]{
  \smallskip \noindent\colorbox{gray!30}{\parbox{\dimexpr\textwidth-8.5pt}{
    \textbf{Modelling Tip \thetips:}\xspace #1}
  }\smallskip \stepcounter{tips}
}

\usepackage{xspace}
\newcommand{\ie}{\emph{i.e.}\@\xspace}
\newcommand{\eg}{\emph{e.g.}\@\xspace}
\makeatletter
\newcommand*{\etc}{\@ifnextchar{.}{etc}{etc.\@\xspace}}
\makeatother

\usepackage[capitalise, nameinlink, sort]{cleveref}

\crefname{lstlisting}{Listing}{Listings}
\Crefname{lstlisting}{Listing}{Listings}

\usepackage{listings}

\definecolor{solarized@base03}{HTML}{002B36}
\definecolor{solarized@base02}{HTML}{073642}
\definecolor{solarized@base01}{HTML}{586e75}
\definecolor{solarized@base00}{HTML}{657b83}
\definecolor{solarized@base0}{HTML}{839496}
\definecolor{solarized@base1}{HTML}{93a1a1}
\definecolor{solarized@base2}{HTML}{EEE8D5}
\definecolor{solarized@base3}{HTML}{FDF6E3}
\definecolor{solarized@yellow}{HTML}{B58900}
\definecolor{solarized@orange}{HTML}{CB4B16}
\definecolor{solarized@red}{HTML}{DC322F}
\definecolor{solarized@magenta}{HTML}{D33682}
\definecolor{solarized@violet}{HTML}{6C71C4}
\definecolor{solarized@blue}{HTML}{268BD2}
\definecolor{solarized@cyan}{HTML}{2AA198}
\definecolor{solarized@green}{HTML}{859900}

\lstdefinelanguage{bigrapher}{
    keywords={fun, brs, end, sbrs, pbrs, abrs, begin, init, atomic, preds,
      rules, actions, big, ctrl, float, int, react, share, by},
    morekeywords={in, if, param, ctx},
    numberstyle=\tiny\color{solarized@base01},
    keywordstyle=\color{solarized@green},
    stringstyle=\color{solarized@cyan}\ttfamily,
    commentstyle=\color{solarized@base01},
    emphstyle=\color{solarized@red},
    comment=[l]{\#},
}

\lstnewenvironment{bigrapher}[1][] {
  \lstset{
    basicstyle=\scriptsize\ttfamily,
    mathescape=true,
    breaklines=true,
    sensitive=true,
    numbers=left,
    numberstyle=\tiny,
    numbersep=3pt,
    xleftmargin=8pt,
    escapeinside={:}{:},
    showstringspaces=false,
    flexiblecolumns=false,
    frame=tb,
    language=bigrapher,
    #1
  }} {}

\newcommand{\biginline}[1]{\unskip\xspace\lstinline[language=bigrapher,basicstyle=\ttfamily]{#1}}

\usepackage[disable]{todonotes}
\newcommand{\bacom}[1]{\todo[inline, color=blue!40]{BA: #1}}
\newcommand{\mscom}[1]{\todo[inline, color=green!40]{MS: #1}}

\newcounter{tips}
\title{Practical Modelling with Bigraphs}
\author{Blair Archibald}
\orcid{}
\affiliation{\institution{University of Glasgow}
  \country{UK}
}
\email{Blair.Archibald@glasgow.ac.uk}

\author{Muffy Calder}
\orcid{}
\affiliation{\institution{University of Glasgow}
  \country{UK}
}
\email{Muffy.Calder@glasgow.ac.uk}

\author{Michele Sevegnani}
\orcid{}
\affiliation{\institution{University of Glasgow}
  \country{UK}
}
\email{Michele.Sevegnani@glasgow.ac.uk}


\begin{abstract}
Bigraphs are a versatile modelling formalism that allows easy expression of placement and connectivity relations in a graphical format.  System evolution is user defined as a set of rewrite rules. 
This paper presents a practical, yet detailed guide to developing, executing, and reasoning about bigraph models, including recent extensions such as parameterised, instantaneous, prioritised and conditional rules, and probabilistic and stochastic rewriting. 

\end{abstract}


\begin{document}

\maketitle

\section{Introduction}\label{sec:intro}
Bigraphs are a universal modelling formalism for describing systems that evolve in space, time, and connectivity. They were introduced by Milner~\cite{milner_SpaceAndMotionOfCommunicatingAgents:2009}, and have been extended to  directed, stochastic,   sharing, conditional, and probabilistic bigraphs ~\cite{grohmann.ea_DirectedBigraphs:2007,DBLP:journals/entcs/KrivineMT08,sevegnani.ea_BigraphsWithSharing:2015,archibald.ea_ConditionalBigraphs:2020,archibald_ProbabilisticBigraphs:2021}. 
While they have seen use in modelling a range of systems including: mixed-reality games~\cite{benford.ea_Savannah:2016}, network management~\cite{DBLP:journals/scp/CalderKSS14}, wireless communication protocols~\cite{DBLP:journals/fac/CalderS14}, biological processes~\cite{DBLP:journals/entcs/KrivineMT08}, cyber-physical security~\cite{Bashir,alrimawi.ea_AutomatedManagementOfSecurityIncidents:2019}, indoor environments~\cite{DBLP:conf/giscience/WaltonW12}, and sensor systems~\cite{sevegnani.ea_ModellingAndVerificationOfSensors:2018}, they have not yet seen widespread adoption. 
One reason is the early
emphasis on 
    theoretical aspects, including the relationship to specific mathematical categories, and tools for deriving bisimulation congruences that are common in work on process calculi.   Less attention was     given to   bigraphs for system modelling and analysis.
The purpose of this  paper is to provide practical guidance on how to model with bigraphs and   to illustrate some of our extensions to   bigraph theory that enhance modelling in practice.

A bigraph    consists of a set of user defined \emph{entity types} relevant to the domain being modelled, \eg\xspace\biginline{Computer}, \biginline{Person}, \biginline{Room}, \biginline{Cell}, \biginline{Protein}, \dots, which can be related  both {\em spatially} through  nesting, \eg a \biginline{Person} \emph{in} a \biginline{Room}, or and through {\em linking}, \eg communication between    \biginline{Computer}s   in different \biginline{Room}s.
Spatial relations are described by  {\em place graphs} (a forest), with {\em regions} indicating modules, or   adjacent parts of the system;     linking is described by {\em link graphs} (a hypergraph). A bigraph {\em reactive system} (BRS)  consists of   bigraphs  and user-defined {\em rewrite rules} that define how    bigraphs evolve over time.
For example, the rules might express  circumstances under which a  \biginline{Person}  
leaves a   \biginline{Room} or a   
\biginline{Computer} is  connected/disconnected to a
\biginline{Network}.

A core advantage of bigraphs over other modelling formalisms is the diagrammatic notation, backed by an \emph{equivalent} algebraic form, which provides intuitive descriptions of systems without requiring detailed knowledge of mathematical description languages.
The notation is expressive, compared with similar diagrammatic formalisms such as Petri-nets~\cite{petrinets}, as  entity types (including connectivity, placement \etc), their diagrammatic representations,  and the rewrite rules for updating a model, are all  \emph{user} defined. We have found the diagrammatic notation particularly useful and accessible to system designers and users when developing     models.

\begin{figure}
  \begin{subfigure}[b]{\linewidth}
    \centering
  \resizebox{0.6\linewidth}{!}{
    \begin{tikzpicture}[
user/.append style = {draw, circle},
street/.append style = {draw},
room/.append style = {draw},
phone/.append style = {draw},
nme/.append style = {draw},
house/.append style = {draw},
display/.append style = {draw, ellipse},
address/.append style = {draw},
]

\node[display,  label={[inner sep=0.5, name=n1l]north:{\sf\tiny Display}}] (n1) {};
\node[phone, right=0.6 of n1, label={[inner sep=0.5, name=n2l]north:{\sf\tiny Phone}}] (n2) {};
\node[user, right=0.6 of n2, label={[inner sep=0.5, name=n3l]north:{\sf\tiny User}}] (n3) {};
\node[nme, right=0.8 of n3, label={[inner sep=0.5, name=n4l]north:{\sf\tiny Name}}] (n4) {};
\node[street, right=0.6 of n4, label={[inner sep=0.5, name=n6l]north:{\sf\tiny Street}}] (n6) {};
\node[house, right=0.6 of n6, label={[inner sep=0.5, name=n7l]north:{\sf\tiny House}}] (n7) {};
\node[room, fit=(n3)(n3l)(n2)(n2l)(n1)(n1l), label={[inner sep=0.5, name=n0l]north:{\sf\tiny Room}}] (n0) {};
\node[address, fit=(n7)(n7l)(n6)(n6l), label={[inner sep=0.5, name=n5l]north:{\sf\tiny Address}}] (n5) {};
\node[big region, fit=(n0)(n0l)] (r0) {};
\node[big region, fit=(n5)(n5l)(n4)(n4l)] (r1) {};
\node[] at ($(r0.north west) + (2.8,0.1)$) (name_r) {\tiny $r$};
\draw[big edge] (n1) to[out=0,in=180] (n2);
\draw[big edge] (n2) to[out=0,in=180] (n3);
\coordinate (h0) at ($(n3) + (0.5,0.3)$);
\draw[big edge] (n3) to[out=0,in=-90] (h0);
\draw[big edge] (n4) to[out=180,in=-90] (h0);
\draw[big edge] (n5) to[out=210,in=-90] (h0);
\draw[big edge] (name_r) to[out=-90,in=90] (h0);

\end{tikzpicture}
   }
  \caption{}
  \end{subfigure}
  
  \begin{subfigure}[b]{0.45\linewidth}
    \centering
  \resizebox{\linewidth}{!}{
    \begin{forest}
  for tree={edge = {-latex}}
  [,phantom 
      [,big region
        [\sf Room
            [\sf Display]
            [\sf Phone]
            [\sf User]
        ]
      ]
      [,big region 
          [ \sf Name  ]
          [ \sf Address 
              [ \sf Street ]
              [ \sf House ]
          ]
      ]
  ]
\end{forest}   }
  \caption{}
  \end{subfigure}
  \begin{subfigure}[b]{0.45\linewidth}
    \centering
  \resizebox{\linewidth}{!}{
    \begin{tikzpicture}[
user/.append style = {draw, circle},
street/.append style = {draw},
room/.append style = {draw},
phone/.append style = {draw},
nme/.append style = {draw},
house/.append style = {draw},
display/.append style = {draw, ellipse},
address/.append style = {draw},
]

\node[display,  label={[inner sep=0.5, name=n1l]north:{\sf\tiny Display}}] (n1) {};
\node[phone, right=0.6 of n1, label={[inner sep=0.5, name=n2l]north:{\sf\tiny Phone}}] (n2) {};
\node[user, right=0.6 of n2, label={[inner sep=0.5, name=n3l]north:{\sf\tiny User}}] (n3) {};
\node[nme, right=0.8 of n3, label={[inner sep=0.5, name=n4l]north:{\sf\tiny Name}}] (n4) {};
\node[street, right=0.6 of n4, label={[inner sep=0.5, name=n6l]north:{\sf\tiny Street}}] (n6) {};
\node[house, right=0.6 of n6, label={[inner sep=0.5, name=n7l]north:{\sf\tiny House}}] (n7) {};
\node[room, above=0.3 of n2l, label={[inner sep=0.5, name=n0l]north:{\sf\tiny Room}}] (n0) {};
\node[address, above=0.3 of n6l, label={[inner sep=0.5, name=n5l]north:{\sf\tiny Address}}] (n5) {};
\node[] at (2.8,1) (name_r) {\tiny $r$};
\draw[big edge] (n1) to[out=0,in=180] (n2);
\draw[big edge] (n2) to[out=0,in=180] (n3);
\coordinate (h0) at ($(n3) + (0.5,0.3)$);
\draw[big edge] (n3) to[out=0,in=-90] (h0);
\draw[big edge] (n4) to[out=180,in=-90] (h0);
\draw[big edge] (n5) to[out=180,in=-90,looseness=0.6] (h0);
\draw[big edge] (name_r) to[out=-90,in=90] (h0);

\end{tikzpicture}
   }
  \caption{}
  \end{subfigure}

  \begin{subfigure}[b]{\linewidth}
    \centering
  \resizebox{0.8\linewidth}{!}{
    
\begin{tikzpicture}[
,
user/.append style = {draw, circle},
room/.append style = {draw},
phone/.append style = {draw},
display/.append style = {draw, ellipse},
]
\begin{scope}[local bounding box=lhs, shift={(0,0)}]

\node[display,  label={[inner sep=0.5, name=n1l]north:{\sf\tiny Display}}] (n1) {};
\node[phone, right=0.60 of n1, label={[inner sep=0.5, name=n2l]north:{\sf\tiny Phone}}] (n2) {};
\node[user, right=0.60 of n2, label={[inner sep=0.5, name=n3l]north:{\sf\tiny User}}] (n3) {};
\node[big site, right=0.60 of n3,] (s0){};
\node[room, fit=(n3)(n3l)(n2)(n2l)(n1)(n1l)(s0), label={[inner sep=0.5, name=n0l]north:{\sf\tiny Room}}] (n0) {};
\node[big region, fit=(n0)(n0l)] (r0) {};
\node[] at ($(r0.north west) + (1.2,0.1)$) (name_w) {\tiny $w$};
\node[right=1.1 of name_w] (name_u) {\tiny $u$};
\draw[big edge] (n2) to[out=0,in=180] (n3);
\draw[big edge] (n3) to[out=0,in=-90] (name_u);
\coordinate (h0) at ($(n1) + (0.3,0.1)$);
\draw[big edge] (n1) to[out=0,in=-90] (h0);
\draw[big edge] (n2) to[out=180,in=-90] (h0);
\draw[big edge] (name_w) to[out=-90,in=90] (h0);

\end{scope}
\begin{scope}[local bounding box=rhs, shift={(4.5,0)}]

\node[display,  label={[inner sep=0.5, name=n1l]north:{\sf\tiny Display}}] (n1) {};
\node[big site, right=0.60 of n1,] (s0){};
\node[phone, right=0.60 of s0, label={[inner sep=0.5, name=n2l]north:{\sf\tiny Phone}}] (n2) {};
\node[user, right=0.60 of n2, label={[inner sep=0.5, name=n3l]north:{\sf\tiny User}}] (n3) {};
\node[room, fit=(n1)(n1l)(s0), label={[inner sep=0.5, name=n0l]north:{\sf\tiny Room}}] (n0) {};
\node[big region, fit=(n3)(n3l)(n2)(n2l)(n0)(n0l)] (r0) {};
\node[] at ($(r0.north west) + (1.5,0.1)$) (name_w) {\tiny $w$};
\node[right=1 of name_w, xshift=1.0] (name_u) {\tiny $u$};
\draw[big edgec] (n2) to[out=180,in=-90] ($(n2) + (-0.4,0.3)$);
\draw[big edge] (n2) to[out=0,in=-90] (n3);
\draw[big edge] (n3) to[out=180,in=-90] (name_u);
\draw[big edge] (n1) to[out=0,in=-90] (name_w);

\end{scope}

\node[xshift=28] at ($(lhs)!0.5!(rhs)$) {$\rrul$};
\end{tikzpicture}

   }
  \caption{}
  \end{subfigure}

  \begin{subfigure}[b]{\linewidth}
    \centering
  \resizebox{\linewidth}{!}{
    
\begin{tikzpicture}[
  ,
user/.append style = {draw, circle},
street/.append style = {draw},
room/.append style = {draw},
phone/.append style = {draw},
nme/.append style = {draw},
house/.append style = {draw},
display/.append style = {draw, ellipse},
address/.append style = {draw}
  ]
  \begin{scope}[local bounding box=lhs, shift={(0,0)}]
\node[display,  label={[inner sep=0.5, name=n1l]north:{\sf\tiny Display}}] (n1) {};
\node[phone, right=0.6 of n1, label={[inner sep=0.5, name=n2l]north:{\sf\tiny Phone}}] (n2) {};
\node[user, right=0.6 of n2, label={[inner sep=0.5, name=n3l]north:{\sf\tiny User}}] (n3) {};
\node[nme, right=0.8 of n3, label={[inner sep=0.5, name=n4l]north:{\sf\tiny Name}}] (n4) {};
\node[street, right=0.6 of n4, label={[inner sep=0.5, name=n6l]north:{\sf\tiny Street}}] (n6) {};
\node[house, right=0.6 of n6, label={[inner sep=0.5, name=n7l]north:{\sf\tiny House}}] (n7) {};
\node[room, fit=(n3)(n3l)(n2)(n2l)(n1)(n1l), label={[inner sep=0.5, name=n0l]north:{\sf\tiny Room}}] (n0) {};
\node[address, fit=(n7)(n7l)(n6)(n6l), label={[inner sep=0.5, name=n5l]north:{\sf\tiny Address}}] (n5) {};
\node[big region, fit=(n0)(n0l)] (r0) {};
\node[big region, fit=(n5)(n5l)(n4)(n4l)] (r1) {};
\node[] at ($(r0.north west) + (2.8,0.1)$) (name_r) {\tiny $r$};
\draw[big edge] (n1) to[out=0,in=180] (n2);
\draw[big edge] (n2) to[out=0,in=180] (n3);
\coordinate (h0) at ($(n3) + (0.5,0.3)$);
\draw[big edge] (n3) to[out=0,in=-90] (h0);
\draw[big edge] (n4) to[out=180,in=-90] (h0);
\draw[big edge] (n5) to[out=210,in=-90] (h0);
\draw[big edge] (name_r) to[out=-90,in=90] (h0);
  \end{scope}
  \begin{scope}[local bounding box=rhs, shift={($(lhs.east) + (1.5,-0.25)$)}]
    
\node[display,  label={[inner sep=0.5, name=n5l]north:{\sf\tiny Display}}] (n5) {};
\node[phone, right=0.60 of n5, label={[inner sep=0.5, name=n6l]north:{\sf\tiny Phone}}] (n6) {};
\node[user, right=0.60 of n6, label={[inner sep=0.5, name=n7l]north:{\sf\tiny User}}] (n7) {};
\node[nme, right=0.60 of n7, label={[inner sep=0.5, name=n0l]north:{\sf\tiny Name}}] (n0) {};
\node[street, right=0.60 of n0, label={[inner sep=0.5, name=n2l]north:{\sf\tiny Street}}] (n2) {};
\node[house, right=0.60 of n2, label={[inner sep=0.5, name=n3l]north:{\sf\tiny House}}] (n3) {};
\node[address, fit=(n3)(n3l)(n2)(n2l), label={[inner sep=0.5, name=n1l]north:{\sf\tiny Address}}] (n1) {};
\node[room, fit=(n5)(n5l), label={[inner sep=0.5, name=n4l]north:{\sf\tiny Room}}] (n4) {};
\node[big region, fit=(n7)(n7l)(n6)(n6l)(n4)(n4l)] (r0) {};
\node[big region, fit=(n1)(n1l)(n0)(n0l)] (r1) {};
\node[] at ($(r0.north west) + (2.7,0.1)$) (name_r) {\tiny $r$};
\draw[big edgec] (n5) to[out=0,in=-90] ($(n5) + (0.3,0.18)$);
\draw[big edgec] (n6) to[out=180,in=-90] ($(n6) + (-0.35,0.3)$);
\draw[big edge] (n6) to[out=0,in=180] (n7);
\coordinate (h0) at ($(n0) + (-0.45,0.3)$);
\draw[big edge] (n0) to[out=180,in=-90] (h0);
\draw[big edge] (n1) to[out=210,in=-90] (h0);
\draw[big edge] (n7) to[out=0,in=-90] (h0);
\draw[big edge] (name_r) to[out=-90,in=90] (h0);

  \end{scope}

  \node[xshift=0] at ($(lhs.east)!0.5!(rhs.west)$) {$\react$};
\end{tikzpicture}
  
   }
  \caption{}
  \end{subfigure}
 
  \caption{Example bigraph with two regions: a physical \biginline{Room} containing a Wi-Fi-enabled \biginline{Display}, a \biginline{User}, and their \biginline{Phone}, and  data consisting of a \biginline{Name} and an \biginline{Address}.  
  (a) Diagrammatic representation: entities are black shapes and links are green lines. 
  (b) The place graph.
  (c) The link graph. 
  (d) Example reaction rule: a user leaves a room and takes the phone with them.
  (e) Application of the reaction rule from (d) to (a).
  }
   \label{fig:intro}
\end{figure}
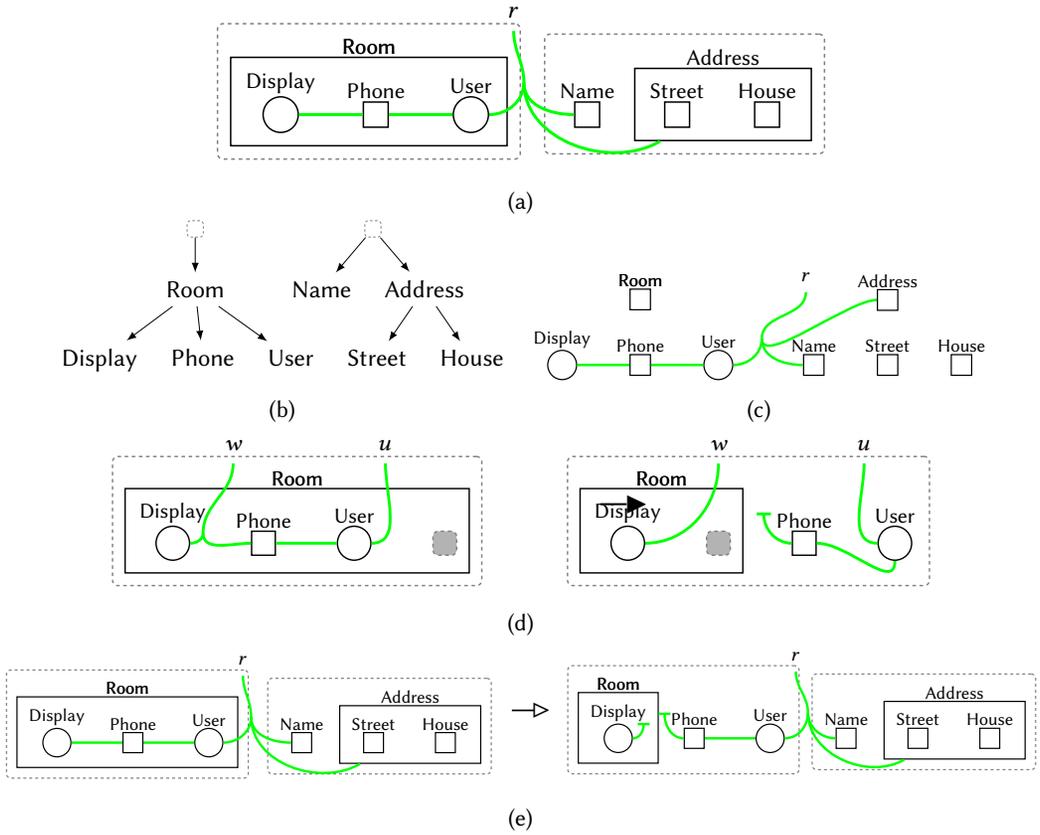

The example   in Fig.~\ref{fig:intro} illustrates   
  a bigraph model, in diagrammatic format. In \cref{fig:intro}a there are two regions
    (the dashed rectangles), indicating there are two distinct parts of the model: physical and data.
  The physical region consists of a \biginline{Room},
  containing a Wi-Fi-enabled \biginline{Display}, a    \biginline{User}, and their \biginline{Phone}, which is connected to the \biginline{Display}. The   data region   consists  of the User's  \biginline{Name}  and  \biginline{Address}, which consists of a \biginline{Street} and
\biginline{House}. In general, we may use (coloured) shapes to denote different entities. Links are in green and may be named, \eg\ {\em r}, {\em w}, which indicates a potential link to other entities.
\cref{fig:intro}d  contains  an example rewrite rule:    a \biginline{User} leaves a \biginline{Room},  taking their \biginline{Phone} with them.    
The \biginline{Display} may be connected to other devices  (\eg more phones or computers), which is indicated,   on the left-hand side, by  the link named {\em w}.    On the right-hand side, the \biginline{User} and their \biginline{Phone} are no longer in the \biginline{Room} and   the \biginline{Display} and \biginline{Phone}
are disconnected, but any other connections {\em w} remain (including no connection).
The gray rectangles denote that anything else that may have been in the \biginline{Room} will remain in the \biginline{Room} unchanged.
\cref{fig:intro}e shows the transition that results when we apply this rewrite rule to our example bigraph. Notice that the data region   is unchanged.

Our   aim is to provide a core intuition behind bigraphs and familiarity with the    diagrammatic and text formats.
To this end, we avoid giving formal mathematical descriptions where possible and focus on example-driven explanations.
We do not assume any existing knowledge, and 
by the end of this paper a reader should be able to create their own bigraph models and use them for rigorous system design and analysis. 
A previous example-driven tutorial~\cite{debois.ea_BigraphsByExample:2005} showed the versatility of bigraphs for   modelling, however it does not account for recent innovations, \eg parameters,  priorities, and conditional bigraphs~\cite{archibald.ea_ConditionalBigraphs:2020}, nor give  detailed descriptions of modelling techniques/styles that are common in practice, \eg how to apply a rewrite rule a certain number of times, or how to model multiple perspectives. Likewise, existing bigraph publications usually show models without explaining in detail why these modelling decisions were made.

To show bigraphs are not just a theoretical tool, but a practical one, we provide fully executable model files of each example~\cite{modelFiles} in BigraphER format, and a wider collection of examples is available online\footnote{\url{https://uog-bigraph.bitbucket.io/examples.html}}.
BigraphER~\cite{sevegnani_BigraphER:2016} is an open-source framework for manipulating, visualising, and executing bigraphical reactive systems by applying rewrites. 
We choose BigraphER as it is actively maintained, features an intuitive syntax similar to the algebraic definition of bigraphs, and has a range of state-of-the-art extensions such as probabilistic rewriting~\cite{archibald_ProbabilisticBigraphs:2021} and conditionals~\cite{archibald.ea_ConditionalBigraphs:2020}.
Examples not requiring these features are applicable to other bigraph tools such as JLibBig~\cite{jlibbig}, which also supports directed bigraphs (we discuss emulating directed links in \cref{sec:dir_links}), or the Bigraph Toolkit Suite~\cite{grzelak_BTS:2021}. Historical tools such as BPL~\cite{hojsgaard.ea_BPLTool:2011} (Bigraphical Programming Language), BigMC~\cite{perrone.ea_BigMC:2012} (Bigraph Model Checker), and BigRed~\cite{faithfull_BigRed:2013} are unmaintained.

\subsection{Paper organisation}

The next five sections cover the basics of bigraphs and BigraphER. 
In \cref{sec:place_graphs} we show how place graphs are used to model topological    relationships, \eg the spatial arrangement of entities;  in   \cref{sec:link_graphs} 
we show how link graphs are used to model linking  relationships such as communication; and in 
 \cref{sec:brs}, we show to model bigraph evolution through (user-defined) rewrite rules.
 \cref{sec:extension1} introduces
 a number of extensions we have defined and implemented: parameterised entities and rules, and instantaneous and    conditional   rules.

The following three sections offer practical modelling advice for common scenarios: multi-perspective modelling in 
 \cref{sec:multiperspective}, entity versus link structures in 
  \cref{sec:practical}, and
  rewriting control    in 
  \cref{sec:tagging},
including
  phases and turn taking.
 
 Further  rewriting extensions:    
  probabilistic, stochastic, and non-deterministic rewriting, are covered in \cref{sec:BRS2}, and in  \cref{sec:analysis} we give an overview of analysis through  state space exploration both using simulation (single trace) and model-checking (sets of traces). We also introduce    bigraph patterns and   give pointers to several detailed bigraph models and their analysis.  

  Modelling tips are given throughout, and summarised in \cref{tips}; we conclude in \cref{sec:conclusions}.

\section{Place Graphs}\label{sec:place_graphs}
The place graph describes the  \emph{topological}  relationships between entities. It is   used to model    spaces, for example location: an  \biginline{Adult}  is in a \biginline{Room}, or ownership: a \biginline{BatteryLevel}   belongs to  a specific \biginline{Phone}.
In standard bigraphs, these relations are  described by \emph{forests} that are a collection of \emph{trees}, \ie you are allowed multiple roots.
In \cref{sec:sharing} we consider an extension     that
allows  directed acyclic graphs instead of forests.

\begin{figure}
  \centering
	\begin{subfigure}[b]{0.4\linewidth}
		\centering
\resizebox{\linewidth}{!}{
      \begin{tikzpicture}[]
  \node[inner sep=0.2] (p1) {\Ladiesroom};
  \node[right=0.01 of p1, inner sep=0.2] (c1) {\resizebox{0.1cm}{!}\Gentsroom};
  \node[room, fit=(p1)(c1), label={[name=r1l]}] (r1) {};
  \node[room, right=0.3 of c1, label={[name=r2l]}] (r2) {};
  \node[room, right=1 of c1, label={[name=r3l]}] (r3) {};
  \node[floor, fit=(r1)(r1l)(r2)(r2l), inner sep=4.5, label={[name=f1l]}] (f1) {};
  \node[floor, fit=(r3)(r3l), inner sep=4.5, label={[name=f2l]}] (f2) {};
  \node[building, fit=(f1)(f1l)(f2)(f2l), label={[name=b1l]}] (b1) {};

  \node[inner sep=0.2, right=2.25 of c1] (p2) {\Ladiesroom};
  \node[room, fit=(p2), label={[name=r4l]}] (r4) {};
  \node[floor, fit=(r4)(r4l), label={[name=f2l]}] (f2) {};
  \node[building, fit=(f2)(f2l), label={[name=b2l]}] (b2) {};

  \node[big region, fit=(b1)(b1l)] (r1) {};
  \node[big region, fit=(b2)(b2l)] (r2) {};
\end{tikzpicture}
     }
		\caption{}
    \label{fig:pg_example_1_a}
	\end{subfigure}
	\begin{subfigure}[b]{0.5\linewidth}
		\centering
    \resizebox{\linewidth}{!}{
      \begin{forest}
  for tree={edge = {-latex}}
  [, phantom
    [0, big region
        [\sf Building, name=b1
        [\sf Floor
        [\sf Room
        [\Ladiesroom, name=p1]
        [\resizebox{0.1cm}{!}{\Gentsroom}, name=p2]
        ]
        [\sf Room]
        ]
        [\sf Floor
            [\sf Room]
        ]]
    ]
    [1, big region
    [\sf Building, name=b2
    [\sf Floor
    [\sf Room
    [\Ladiesroom
    ]]]]]
  ]
  \node[draw, rounded rectangle, dashed, red, thick, fit=(b1)(b2)] (par) {};
  \node[red] at ($(b1)!0.5!(b2)$) {\biginline{||}};
  \node[draw, rounded rectangle, dashed, blue, thick, fit=(p1)(p2)] (par) {};
  \node[blue] at ($(p1)!0.5!(p2)$) {\biginline{|}};
\end{forest}
     }
		\caption{}
    \label{fig:pg_example_1_b}
	\end{subfigure}
	\begin{subfigure}[b]{\linewidth}
		\centering
    \begin{bigrapher}
atomic ctrl Adult = 0;
atomic ctrl Child = 0;
ctrl Building = 0;
ctrl Floor = 0;
ctrl Room = 0;

big space =
     Building.(Floor.(Room.(Adult | Child) | Room.1) | Floor.Room.1)
  || Building.Floor.Room.Adult;
    \end{bigrapher}
		\caption{}
    \label{fig:pg_example_1_c}
	\end{subfigure}
  \caption{Modelling   buildings. (a) Diagrammatic   place graph (b) Forest representation (c) BigraphER model. In (b),      the red and blue dashed ovals, containing   red and blue parallel and merge product operators resp., are superposed on the place graph. These are not part of bigraph notation, but serve to highlight the difference between \biginline{||} and \biginline{|}.\mscom{not entirely happy with the icons for Adult and child\bacom{Think we would need to just use labelled shapes otherwise? But can do so if you want}\mscom{I mainly don't like the skirt but let's wait for Muffy's opinion}}  }
  \label{fig:pg_example_1}
\end{figure}
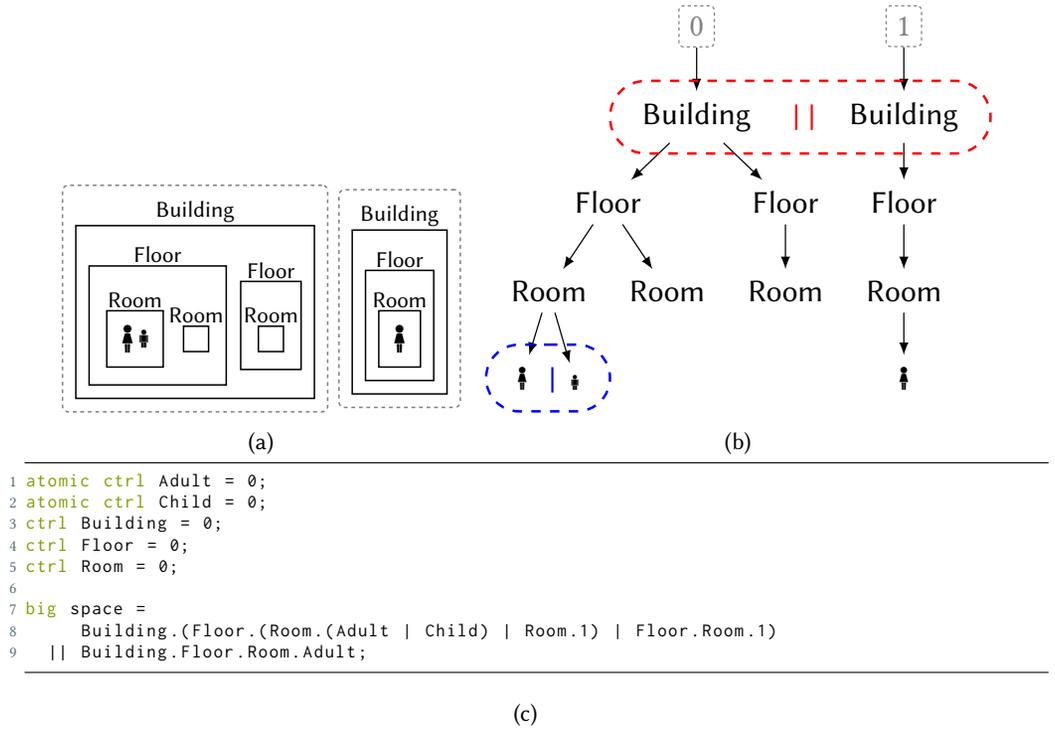

To illustrate place graphs, we show a model of   locations and people (adults and children) within a building in \cref{fig:pg_example_1}.
We give three equivalent representations: \cref{fig:pg_example_1_a}
shows the standard way to draw a place graph where \emph{nesting} is explicit, \cref{fig:pg_example_1_b} shows the equivalent place graph as the underlying forest, and \cref{fig:pg_example_1_c} shows the BigraphER code corresponding to this graph.
The binary operator \biginline{.} is used to indicate nesting.   For example, 
 \biginline{Room.Adult} means an \biginline{Adult} within a \biginline{Room}.
 The keyword \biginline{atomic} indicates  an entity that cannot contain any children.

Types are defined with keyword \biginline{ctrl} \footnote{We prefer the reserved word {\em type} but have remained faithful to Milner's reference to {\em control}.}. Keyword  \biginline{big} 
introduces a named bigraph.
BigraphER comments are indicated by \#. 
Entities are allowed any number of children---including none which is represented by the special bigraph  \biginline{1}---and siblings are defined using \emph{merge product} written \biginline{|}.
Note that each entity also has 
``\biginline{= 0}''. We have included this so that the BigraphER code is correctly formed; strictly, this refers to the link graph, which will be explained in \cref{sec:link_graphs}.

For example, \biginline{Room.(Adult | Child)} is a room containing one adult and one child.
Merge product is \emph{commutative}, meaning children are \emph{unordered}.
That is, \biginline{Room.(Adult | Child)} and \biginline{Room.(Child | Adult)}
model the same room; 
if ordering is required it must be explicitly encoded (see \cref{sec:ordered_children}).

To create multiple \emph{regions} (roots) we have the operator \biginline{||} called \emph{parallel product}.
For example, \biginline{Building || Building} is the forest with two trees, each containing a building. 

\tipbox{\biginline{||} and \biginline{|} allow us to build bigger bigraphs from smaller.  Use \biginline{||} to model distinct bigraphs and \biginline{|} for merging bigraphs.}

Notice that entities are typed, but do not have  names,  \eg       there may be several  \biginline{Room} entities in a \biginline{Floor} but   we cannot identify a specific  one. Formally, bigraphs without identifiers are called
\emph{abstract bigraphs} whereas
\emph{concrete bigraphs} assign identifiers to entities.
In this paper we refer exclusively to abstract bigraphs.

\subsection{Regions and sites}
Bigraphs are \emph{always} rooted\footnote{This only applies to standard bigraphs. Bigraphs with sharing allows bigraphs with no parents.} using \emph{regions}, represented by the dashed rectangles. Regions  indicate adjacent parts of the system. In the buildings example there are two regions, one for each building.  

We stated above that the   
  place graph \one\; indicates an entity has no children; now we can be more precise---it   represents the place graph with a single region and no nothing else.

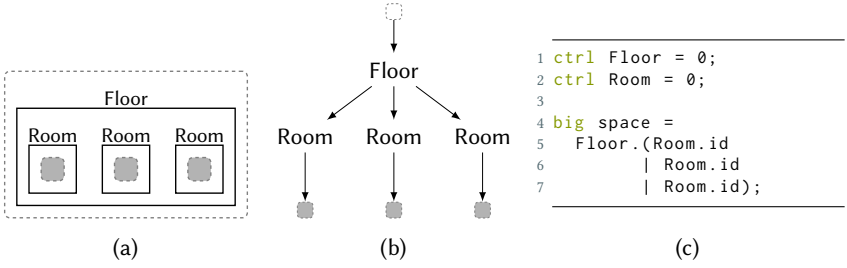
\begin{figure}
  \centering
	\begin{subfigure}[b]{0.25\linewidth}
		\centering
    \resizebox{\linewidth}{!}{
      \begin{tikzpicture}[]
  \node[big site] (s1) {};
  \node[big site, right=0.5 of s1] (s2) {};
  \node[big site, right=0.5 of s2] (s3) {};
  \node[room, fit=(s1), label={[name=r1l]}] (r1) {};
  \node[room, fit=(s2), label={[name=r2l]}] (r2) {};
  \node[room, fit=(s3), label={[name=r3l]}] (r3) {};
  \node[floor, fit=(r1)(r1l)(r2)(r2l)(r3)(r3l), label={[name=f1l]}] (f1) {};
  \node[big region, fit=(f1)(f1l)] (reg1) {};
\end{tikzpicture}
     }
		\caption{}
    \label{fig:pg_example_2_a}
	\end{subfigure}
	\begin{subfigure}[b]{0.25\linewidth}
		\centering
    \resizebox{\linewidth}{!}{
      \begin{forest}
  for tree={edge = {-latex}}
  [,big region
    [\sf Floor
        [\sf Room [,big site]]
        [\sf Room [,big site]]
        [\sf Room [,big site]]]]
\end{forest}
     }
		\caption{}
    \label{fig:pg_example_2_b}
	\end{subfigure}
	\begin{subfigure}[b]{0.3\linewidth}
		\centering
    \begin{bigrapher}
ctrl Floor = 0;
ctrl Room = 0;

big space =
  Floor.(Room.id
        | Room.id
        | Room.id);
    \end{bigrapher}
		\caption{}
    \label{fig:pg_example_2_c}
	\end{subfigure}
  \caption{Place graph with one region and three  sites.}
  \label{fig:pg_example_2}
\end{figure}

We can abstract away from entities using  \emph{sites}, which are like variables (see \cref{sec:instantiation}). We draw them as dashed gray filled rectangles.  For example, in \cref{fig:pg_example_2}, each \biginline{Room} contains one site that represents one or more bigraphs. 
 A site can be nested wherever an entity can be nested, including directly under a region, in which case  this region/site pair is called the \emph{identity} place graph    and denoted by \biginline{id}. 
 
 An example containing   regions and sites is in \cref{fig:pg_example_2}, where we focus on a single floor of a building,  having abstracted away the specific contents of rooms using sites.

\subsection{Two special place graphs: \biginline{id} and \biginline{1}}
     
 We  use   \one\  with a {\em non-atomic} entity to indicate that sites  are not required, for example, when a room is empty, as indicated in line 8 of \cref{fig:pg_example_1_c}.
 Notice that in the diagrammatic form, \eg 
 in \cref{fig:pg_example_1_a}, we simply do not draw anything underneath \biginline{Room} to indicate an empty room, whereas in the textual form, we have to make this explicit  with   the  text \one\ (as an operand of the nesting operator \biginline{.}).  
 
 To summarise the difference between \biginline{1} and \biginline{id},  consider 
Fig.'s~\ref{fig:pg_example_1} and~\ref{fig:pg_example_2}. 
\biginline{Room.1} indicates a room  with no possibility of children, whereas \biginline{Room.id} indicates a room with a site, which may be instantiated with \emph{zero} or more (children). That is, sites might themselves contain \one, meaning there is nothing inside them.
 
\tipbox{Use \biginline{1}  to indicate  ``no possibility of any children''  and  \biginline{id} 
to indicate ``zero or more children''.  }
 
 \subsection{Diagrammatic notation}
The shapes and colours of entities in the diagrammatic\footnote{This is sometimes referred to as graphical notation, which we  avoid  in case of confusion  with place and link {\em graphs.}} notation are chosen by the user.
In this paper we mainly use simple geometric shapes that are easy to draw such as square, rectangle, and circle, but any shape is possible, as well as colour or shading.

Shading might be used to represent a combination of bigraphs that denotes an entity in different states or stages of a process, \eg we might use a circle for \biginline{File} and a red circle for \biginline{File.Open} or alternatively to represent a new entity type \biginline{FileOpen}.
Note that in the theory there is no subtyping so \biginline{File} and \biginline{FileOpen} are distinct entities. 
By giving them the same shape we are giving (informally) an additional relationship between these entities.

\tipbox{Use prefixes/suffixes in entity names (in textual format) and colours and shading (in diagrammatic format) to indicate relationships between states or stages of a process.  }

\subsection{Sharing}
\label{sec:sharing}

In the standard definition of bigraphs, each entity is  allowed only a single parent (another entity or a region).
However,  it may be natural to  allow a single entity to have   multiple parents, for example when modelling spatial overlap  such as wireless signal ranges or fields of vision.     
 
  \emph{Bigraphs with Sharing}~\cite{sevegnani.ea_BigraphsWithSharing:2015} relax  the restriction on place graphs from a forest to a directed acyclic graph, meaning entities can have any number of parents (including being in multiple regions; or having no parents at all).
While there are a few key theoretical implications,
from
  a modelling viewpoint, bigraphs with sharing are a simple extension.

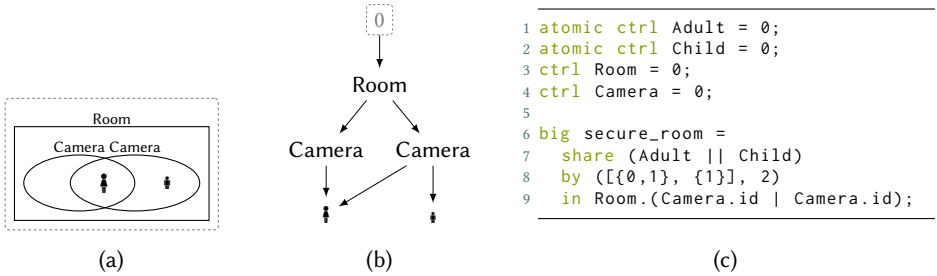
\begin{figure}
  \centering
	\begin{subfigure}[b]{0.25\linewidth}
		\centering
    \begin{tikzpicture}
  \node[inner sep=0.2] (p1) {\Ladiesroom};
  \node[inner sep=0.2, right=0.7 of p1] (p2) {\resizebox{0.1cm}{!}\Gentsroom};
\node[inner sep=0.2, left=0.5 of p1, opacity=0] (spacer) {\Ladiesroom};
  \node[camera, fit=(p1)(spacer), label={[name=c1l]}] (c1) {};
  \node[camera, fit=(p1)(p2),label={[name=c2l]}] (c2) {};
  \node[room, fit=(c1)(c1l)(c2)(c2l), label={[name=rm1l]}] (rm1) {};
  \node[big region, fit=(rm1)(rm1l)] (r1) {};
\end{tikzpicture}
 \caption{}
    \label{fig:pg_sharing_1}
	\end{subfigure}
	\begin{subfigure}[b]{0.25\linewidth}
		\centering
    \resizebox{0.8\linewidth}{!}{
      \begin{forest}
  for tree={edge = {-latex}}
  [0, big region
  [\sf Room
  [\sf Camera, name=c1
  [\Ladiesroom, name=p1]]
  [\sf Camera, name=c2
  [\resizebox{0.1cm}{!}{\Gentsroom}, name=p2]
  ]]]
  \draw[-latex] (c2) -- (p1);
\end{forest}
     }
\caption{}
    \label{fig:pg_sharing_2}
	\end{subfigure}
	\begin{subfigure}[b]{0.4\linewidth}
		\centering
    \begin{bigrapher}
atomic ctrl Adult = 0;
atomic ctrl Child = 0;
ctrl Room = 0;
ctrl Camera = 0;

big secure_room =
  share (Adult || Child)
  by ([{0,1}, {1}], 2)
  in Room.(Camera.id | Camera.id);
    \end{bigrapher}
		\caption{}
    \label{fig:pg_sharing_3}
	\end{subfigure}
  \caption{Bigraphs with sharing:  cameras with overlapping fields of vision that may capture a single \biginline{Adult}. 
  (a) Bigraph model.
  (b) Place graph.
  (c) BigraphER snippet.
  }   
  
  \label{fig:sharing}
\end{figure}

We illustrate by      introducing security  \biginline{Cameras} into our building model.  
A room may have multiple security cameras, each of whose field of vision might overlap, meaning that a single \biginline{Adult} entity might be nested under two (or more) cameras at a time. This is shown
in \cref{fig:pg_sharing_3}, where
line 8   \biginline{by ([\{0,1\}, \{1\}], 2)}  indicates
the first region (containing \biginline{Adult})
should appear in both sites \biginline{\{0,1\}} while the second region 
(containing \biginline{Child})
should only appear in site \biginline{\{1\}}.
The additional \biginline{2} is required as the mapping might not be surjective, \eg we may choose to ignore a region of the \biginline{share} bigraph.

\section{Link Graphs} \label{sec:link_graphs}

\begin{figure}
  \centering
	\begin{subfigure}[b]{0.4\linewidth}
		\centering
    \resizebox{0.6\linewidth}{!}{
      \begin{tikzpicture}
  \node[camera] (c1) {};
  \node[camera, right=0.4 of c1] (c2) {};
  \node[camera, right=0.4 of c2] (c3) {};
  \node[ctrlPanel, below=0.3 of c2] (ctrl1) {};
  \node[below=0.6 of c1, inner sep=0.1] (p1) {\Ladiesroom};
  \node[right=0.3 of p1, inner sep=0.1] (p2) {\Ladiesroom};
  \node[right=0.3 of p2, inner sep=0.1] (p4) {\Ladiesroom};
  \node[right=0.3 of p4, inner sep=0.1] (child) {\resizebox{0.1cm}{!}\Gentsroom};
  \node[right=0.3 of ctrl1, inner sep=0.1] (p3) {\Ladiesroom};

  \coordinate (h1) at ($(c2.south) + (0,-0.2)$);
  \draw[big edge] (c1.south) to[out=-90, in=90] (h1);
  \draw[big edge] (c2.south) to[out=-90, in=90] (h1);
  \draw[big edge] (c3.south) to[out=-90, in=90] (h1);
  \draw[big edge] (h1) to[out=-90, in=0] (ctrl1.north);
\node[ctrlPanel, below=0.3 of c2] (ctrl1) {};

  \draw[big edge] (p1.east) to[out=0, in=180] (p2.west);
  \draw[big edge] (p3.west) to[out=180, in=0] (ctrl1.east);
  \draw[big edgec] (p4.east) to[out=0, in=-90] ($(p4.east) + (0.1,0.1)$);

\end{tikzpicture}
     }
\caption{}
    \label{fig:lg_example_1}
	\end{subfigure}
	\begin{subfigure}[b]{0.5\linewidth}
		\centering
    \begin{bigrapher}
ctrl Camera = 1;
atomic ctrl CtrlPanel = 2;
atomic ctrl Adult = 1;
atomic ctrl Child = 0;

big comms = /x/y/z (
 Camera{x}.1 | Camera{x}.1 | Camera{x}.1
 | CtrlPanel{x,y} | Adult{y}
 | Adult{z} | Adult{z}
 | /c Adult{c} | Child
 );
    \end{bigrapher}
		\caption{}
    \label{fig:lg_example_1_big}
	\end{subfigure}
  \caption{Link graph example: CCTV, phone calls, and remote network access. (a) Link graph. (b) BigraphER snippet.}
  \label{fig:lg_example}
\end{figure}

Bigraphs   allow us to express non-spatial connectivity through the \emph{link graph}, which  is a set of \emph{hyperedges}.  Each hyperedge consists of a non-empty set of vertices.
This contrasts with conventional graph edges  that define  one-to-one relationships (\ie a hypergraph where all sets have cardinality two).

  Each entity involved has a  \textbf{fixed} \emph{arity} that defines the number of \emph{ports}: a name that reflects their role in communication. In the diagrammatic form, ports are implicit---the point at which the edge meets the entity boundary.
  In the text form, \ie BigraphER snippet, 
entity definitions include their arities following the symbol \biginline{=}.  For example  \eg\xspace \biginline{atomic ctrl Camera = 1}, specifies a \biginline{Camera} entity has 1 port.
   Much like the children of an entity in the place graph, \textbf{the ports of an entity are unordered.}
 
  Intuitively, this means you can
  say ``there is \emph{a} port that connects to an   entity'' but not ``port 0 connects to an   entity''.
We show how this can be overcome, if needed, in \cref{sec:typed_ports}.

A link may be 
{\em disconnected}, \ie the entity is not linked to any other entity, (a one-to-zero hyperedge---a singleton set). This is drawn as a single edge with a short perpendicular line at one end (not the entity).

To illustrate link graphs, in \cref{fig:lg_example_1} we have security cameras that can be linked to form CCTV, and adults that can communicate    via phones  or      dial into the  security control panel. Links are drawn in green (any colour can be used). 
All \biginline{Camera} entities have arity 1 and are connected via a single hyperlink to the security control panel.
\biginline{Adult} entities also have arity 1 allowing them to communicate, either with other people or via remote access to the  security control panel.  
One of the  \biginline{Adult}s (third from left)  is  not connected anywhere and so their link  disconnected;  the  single \biginline{Child} entity   has no ports. Note, all 
 entities appear in the link graph, regardless of whether they have any links. 
 
 \subsection{Closed and open links}\label{sec:named}
 
A link may be partially specified, in which case it is  {\em named} to indicate it {\em may}  have connections elsewhere.  An example of this is the link named {\em w} in \cref{fig:intro}, which indicates   the display may link to other devices.
This use of a name is similar to a free variable and we call such a link {\em open}.
The alternative is a {\em closed} link, which is fully specified, \eg the link in \cref{fig:intro}  between the  \biginline{Phone} and \biginline{User}.

In the text form of bigraphs we use identifiers \eg $x$, $y$ to define both links and names, the idea being that ports that share an identifier share a link. An example is in BigraphER code   of \cref{fig:lg_example_1_big}.   
The    {\em closure} operator \biginline{/} indicates   
the  link  that is identified is closed.   
For example,  
 \eg the link identified by \biginline{x} is bound in \biginline{/x (Device\{x\} | Device \{x\})} and so the link is closed and cannot connect elsewhere.   
 Like bound variables in programming languages, the actual identifier does not matter, and so \biginline{/y (Device\{y\}  | Device\{y\})} is an equivalent bigraph.  This also means \textbf{there is no \emph{global} way to refer to a specific link}, \ie identifiers do not identify specific edges and so you cannot refer to, for example, ``the edge named $y$''.
 When an identifier is not bound it becomes a \emph{name}, allowing the link it defines to be extended.
 Finally, a name is \emph{idle} when it exists but is not connected to any other entities or names.
 
 We illustrate links and names in   \cref{fig:names}, which is  based on the declaration  
\biginline{atomic ctrl Device = 1} that indicates 
a device   has no children and one port.  \mscom{Wondering if something like \biginline{/\{x,y\} B\{x,y\}} ond/or \biginline{z/\{x,y\} B\{x,y\}} would be useful as example but we haven't introduced renamings, identities, etc yet}
\bacom{I don't think we use often enough in practice, probably more confusion than benefits?}

\begin{table}
	\centering
 \caption{Links and names: BigraphER notation on the left, diagrammatic notation on the right
 (a) A closed link between two connected devices, no other connections are possible.
 (b) Two connected devices with a further potential connection named $x$.
 (c) A disconnected device.
 (d) A device with a potential link $x$.
 (e) A closed link (to the left) and a open link (to the right),  $x$ is bound in the former and free in the latter.
 (f) An idle link $x$.}
    \footnotesize
	\begin{tabular}{@{}lcc@{}}
      (a) & \biginline{/x (Device \{x\} | Device\{x\})} &
    \resizebox{0.2\textwidth}{!}{
     \begin{tikzpicture}
     \node[draw, circle, label={[inner sep=0.5]south:{\sf\tiny Device}}] (d1) {};
     \node[right=0.5 of d1, draw, circle, label={[inner sep=0.5]south:{\sf\tiny Device}}] (d2) {};
     \draw[big edge] (d1) -- (d2);
     \end{tikzpicture}
                                                          }
     \\
      (b) & \biginline{Device\{x\} | Device\{x\}} &
    \resizebox{0.2\textwidth}{!}{
     \begin{tikzpicture}
     \node[draw, circle, label={[inner sep=0.5]south:{\sf\tiny Device}}] (d1) {};
     \node[right=0.5 of d1, draw, circle, label={[inner sep=0.5]south:{\sf\tiny Device}}] (d2) {};
     \node[inner sep=1] at ($($(d1.east)!0.5!(d2.west)$) + (0,0.5)$) (x) {\tiny $x$};
     \coordinate (h1) at ($(d1.east)!0.5!(d2.west)$);
     \draw[big edge] (d1) -- (h1);
     \draw[big edge] (d2) -- (h1);
     \draw[big edge] (h1) -- (x.south);
     \end{tikzpicture}
                                                    }
                                                    \\
      (c) & \biginline{/x Device\{x\}} &
    \resizebox{0.1\textwidth}{!}{
     \begin{tikzpicture}
     \node[draw, circle, label={[inner sep=0.5]south:{\sf\tiny Device}}] (d1) {};
     \draw[big edgec] (d1.north) -- ($(d1.north) + (0,0.2)$);
     \end{tikzpicture}
                                         }
      \\
      (d) & \biginline{Device\{x\}} &
    \resizebox{0.1\textwidth}{!}{
     \begin{tikzpicture}
     \node[draw, circle, label={[inner sep=0.5]south:{\sf\tiny Device}}] (d1) {};
     \node[above=0.1 of d1, inner sep=1] (x) {\tiny $x$};
     \draw[big edge] (d1.north) -- (x.south);
     \end{tikzpicture}
                                      } \\
      (e) & \biginline{/x (Device\{x\} | Device\{x\}) | Device\{x\}} &
    \resizebox{0.25\textwidth}{!}{
     \begin{tikzpicture}
     \node[draw, circle, label={[inner sep=0.5]south:{\sf\tiny Device}}] (d1) {};
     \node[right=0.5 of d1, draw, circle, label={[inner sep=0.5]south:{\sf\tiny Device}}] (d2) {};
     \draw[big edge] (d1) -- (d2);
     \node[right=0.5 of d2, draw, circle, label={[inner sep=0.5]south:{\sf\tiny Device}}] (d3) {};
     \node[above=0.1 of d3, inner sep=1] (x) {\tiny $x$};
     \draw[big edge] (d3.north) -- (x.south);
     \end{tikzpicture}
                                                                       } \\
      (f) & \biginline{\{x\}} &
    \resizebox{0.08\textwidth}{!}{
     \begin{tikzpicture}
     \node[] (x) {\tiny $x$};
     \end{tikzpicture}
     }
    \end{tabular}
    \label{fig:names}
\end{table}

\tipbox{Use a named, open   link   to indicate ``this link \emph{potentially} connects elsewhere'',  and  a  closed link    
to indicate ``\emph{only} these entities are connected''. The specific names used do  not matter.}

\subsection{Bigraphs: Combining Place and Link Graphs}

\begin{table}
	\centering
  \footnotesize
	\caption{Main Bigraph elements and syntax.}
	\begin{tabular}{@{}lcc@{}}
		\textbf{Element} & \textbf{BigraphER Syntax} & \textbf{Diagram} \\
		\toprule
		Entity of arity 1 & \biginline{K\{a\}} &
																					\raisebox{-10pt}{
																									 \begin{tikzpicture}[]
																										 \node[ellipse,draw] (k) {};
																										 \node[anchor=south east, inner sep=0.5] at (k.north west) (k_lbl) {\tiny \sf K};
																										 \node[above=0.3 of k] (a) {$a$};
																										 \draw[big edge] (k.north) to[in=-90] (a.south);
                                                                                                         \node[big region, fit=(k)(k_lbl)] (r1) {};
																									 \end{tikzpicture}
																					} \\
		Name closure & \biginline{/a K\{a\}} &
																					\raisebox{-2pt}{
																									 \begin{tikzpicture}[]
																										 \node[ellipse,draw] (k) {};
																										 \node[anchor=south east, inner sep=0.5] at (k.north west) (k_lbl) {\tiny \sf K};
																										 \node[above=0.3 of k] (a) {};
																										 \draw[big edgec] (k.north) to[in=-90] (a.south);
                                                                                                         \node[big region, fit=(k)(k_lbl)] (r1) {};
																									 \end{tikzpicture}
																					} \\[0.2cm]
		Identity Place Graph & \biginline{id} &
																									 \begin{tikzpicture}[]
																										 \node[big site] (s1) {};
																										 \node[big region, fit=(s1)] (r1) {};
																									 \end{tikzpicture} \\[0.2cm]
		Identity Link Graph & \biginline{id\{x\}} &
                                                \raisebox{-2pt}{
																									 \begin{tikzpicture}[]
																										 \node[] (xo) {$x$};
																										 \node[below=0.2 of xo] (xi) {$x$};
                                                     \draw[big edge] (xo) -- (xi);
																									 \end{tikzpicture}
                                                } \\[0.2cm]
		Empty Region & \biginline{1} &
																									 \begin{tikzpicture}[]
																										 \node[big region] (r1) {};
																									 \end{tikzpicture} \\[0.2cm]
		Idle (outer) name & \biginline{\{x\}} &
																									 \begin{tikzpicture}[]
																										 \node[] (x) {$x$};
																									 \end{tikzpicture} \\
		\midrule
		Nesting & \biginline{/x A\{x\}.B\{x\}.id}  &
																					\raisebox{-10pt}{
																									 \begin{tikzpicture}[]
																										 \node[big site] (s1) {};
																										 \node[ellipse,draw, fit=(s1), minimum width=20] (b) {};
																										 \node[anchor=south east, inner sep=0.5] at (b.north west) (b_lbl) {\tiny \sf B};
																										 \node[ellipse,draw, fit=(b), xshift=-1.2] (a) {};
																										 \node[anchor=south east, inner sep=0.5] at (a.north west) (a_lbl) {\tiny \sf A};
                                                     \draw[big edge] (a.north) -- (b.north);
                                                                                                         \node[big region, fit=(a)(a_lbl)] (r1) {};
																									 \end{tikzpicture}
																									 } \\
		Parallel product  & \biginline{C\{x\}.id || D\{x\}.id} &
																					\raisebox{-10pt}{
																									 \begin{tikzpicture}[]
																										 \node[big site] (s1) {};
																										 \node[ellipse,draw, fit=(s1), minimum width=20] (c) {};
																										 \node[anchor=south east, inner sep=0.5] at (c.north west) (c_lbl) {\tiny \sf C};
																										 \node[big region, fit=(c)(c_lbl)] (r1) {};

																										 \node[big site, right=1.1 of s1] (s2) {};
																										 \node[ellipse,draw, fit=(s2), minimum width=20] (d) {};
																										 \node[anchor=south east, inner sep=0.5] at (d.north west) (d_lbl) {\tiny \sf D};
																										 \node[big region, fit=(d)(d_lbl)] (r2) {};

																										 \coordinate (mid) at ($(r1.east)!0.5!(r2.west)$);
																										 \node[] at ($(mid) + (0,0.6)$) (a) {$x$};

																										 \draw[big edge] (c.east) to[out=0, in=-90] (a);
																										 \draw[big edge] (d.west) to[out=-180, in=-90] (a);
																									 \end{tikzpicture}
																									 } \\
		Merge product  & \biginline{C\{x\}.id | D\{x\}.id} &
																					\raisebox{-10pt}{
																									 \begin{tikzpicture}[]
																										 \node[big site] (s1) {};
																										 \node[ellipse,draw, fit=(s1), minimum width=20] (c) {};
																										 \node[anchor=south east, inner sep=0.5] at (c.north west) (c_lbl) {\tiny \sf C};

																										 \node[big site, right=1.1 of s1] (s2) {};
																										 \node[ellipse,draw, fit=(s2), minimum width=20] (d) {};
																										 \node[anchor=south east, inner sep=0.5] at (d.north west) (d_lbl) {\tiny \sf D};

																										 \node[big region, fit=(c)(c_lbl)(d)(d_lbl)] (r1) {};

																										 \coordinate (mid) at ($(c.east)!0.5!(d.west)$);
																										 \node[] at ($(mid) + (0,0.6)$) (a) {$x$};

																										 \draw[big edge] (c.east) to[out=0, in=-90] (a);
																										 \draw[big edge] (d.west) to[out=-180, in=-90] (a);
																									 \end{tikzpicture}
																									 } \\
	\end{tabular}
	\label{tab:bigraph_elems}
\end{table}

A bigraph consists of a place graph (\cref{sec:place_graphs}) and a link graph (\cref{sec:link_graphs}) defined over the same set of entities.   
A summary of the main components of bigraphs is in \cref{tab:bigraph_elems}. 

An example bigraph bringing together place and link graphs is in \cref{fig:bigraph_example}.
When drawing bigraphs we overlay the link structure on the place structure. By convention, open link names are always drawn above a bigraph. 
Finally, we  introduce additional terminology that   allows us to ensure  correct bigraph composition and rewriting:    
link graphs have {\em inner} and {\em outer} names.  Outer names are the open links.
  Inner names occur rarely  when developing a   bigraph as a model, and do not occur in any examples in this paper; inner names  exist mainly in internal bigraphs during rewriting. We note there is a similar concept for place graphs, but does  not require additional terminology:     {\em  sites} are {\em inner}  and roots  are  
{\em outer}.

\begin{figure}
  \centering
	\begin{subfigure}[b]{\linewidth}
		\centering
\resizebox{0.5\linewidth}{!}{
    \begin{tikzpicture}
\node[inner sep=0.2] (p1) {\Ladiesroom};
  \node[inner sep=0.2, right=0.5 of p1] (p2) {\resizebox{0.1cm}{!}\Gentsroom};
  \node[inner sep=0.2, right=1 of p2] (p3) {\Ladiesroom};
  \node[big site, right=0.5 of p3] (s1) {};
  \node[ctrlPanel, right=1 of s1, label={[name=ctrl1l]}] (ctrl1) {};
  \node[inner sep=0.2,right=0.5 of ctrl1] (p4) {\Ladiesroom};
  \node[big site,right=0.3 of p4] (s2) {};

  \node[camera, fit=(p1)(p2), label={[name=c1l]}] (c1) {};
  \node[camera, fit=(p3), label={[name=c2l]}] (c2) {};

  \node[room, fit=(c1)(c1l), label={[name=r1l]}] (r1) {};
  \node[room, fit=(c2)(c2l), label={[name=r2l]}] (r2) {};
  \node[room, fit=(p4)(ctrl1)(ctrl1l), label={[name=r3l]}] (r3) {};

  \node[floor, fit=(r1)(r1l)(r2)(r2l)(s1), label={[name=f1l]}] (f1) {};
  \node[floor, fit=(r3)(r3l)(s2), label={[name=f2l]}] (f2) {};

  \node[big region, fit=(f1)(f1l)] (reg1) {};
  \node[big region, fit=(f2)(f2l)] (reg1) {};

\draw[big edge] (p1) to[out=-90, in=-90] (p3);
  \draw[big edge] (p4) to[out=180, in=0] (ctrl1);

\coordinate (h1) at ($($(c2)!0.5!(ctrl1)$) + (0.3,1)$);
  \node[] at ($(h1) + (0,0.7)$) (x) {$x$};
  \draw[big edge] (c1.20) to[in=180] (h1);
  \draw[big edge] (c2.20) to[in=180] (h1);
  \draw[big edge] (ctrl1.180) to[out=180,in=0] (h1);
  \draw[big edge] (x.-90) to[out=-90, in=90] (h1);
\end{tikzpicture}
     }
    \caption{Bigraph model of a building.}
  \end{subfigure}

	\begin{subfigure}[b]{0.8\linewidth}
		\centering
    \begin{bigrapher}
atomic ctrl Adult = 1;
atomic ctrl Child = 0;
ctrl Floor = 0;
ctrl Room = 0;
ctrl Camera = 1;
atomic ctrl CtrlPanel = 2;

big space =
    Floor.(/y (Room.(Camera{x}.(Adult{y} | Child))
             | Room.(Camera{x}.Adult{y})) | id)
 || Floor.(Room.(/z (CtrlPanel{x,z} | Adult{z})) | id);
    \end{bigrapher}
		\caption{}
    \label{fig:bigraph_example_bigrapher}
	\end{subfigure}
  \caption{Example bigraph model of a building. (a) Bigraph representation (b) BigraphER snippet. There is one open link named \emph{x}.} 
  \label{fig:bigraph_example}
\end{figure}

\section{Bigraphical Reactive Systems}
\label{sec:brs}
To encode system dynamics, \ie how bigraphs {\em evolve},  we require a set of  reaction rules (also called rewrite rules). A distinctive feature of bigraphs as a modelling tool is that the rules are \emph{user defined}. Bigraphs together with rewrite rules are known as a Bigraphical Reactive System (BRS).  
 
 Reaction rules  
have  form $L \rrul R$, where 
$L$ and  $R$ are bigraphs; the rule specifies that an instance of $L$ can be substituted by $R$.
To apply a rewrite rule, we find an occurrence of sub-bigraph $L$ within a larger bigraph $B$ and substitute $L$ with $R$ creating a new larger bigraph (state) $B'$.
The relation induced by the rewrite rules is denoted by $\react$; \ie $B \react B'$        if  $B$ can rewrite to $B'$ by application of a of rule.  While in general $L$ and $R$ can be any bigraphs, \textbf{the \emph{interfaces} (sites/regions/names) of $L$ and $R$ must be equal.}

When more than one reaction rule applies, or the same reaction rule has multiple occurences, then the rule applies non-deterministically, \ie we pick any and apply it. 
\begin{figure}
  \centering
\begin{subfigure}[b]{0.4\linewidth}
    \resizebox{\linewidth}{!}{
      \begin{tikzpicture}
  \begin{scope}[local bounding box=lhs]
    \node[inner sep=0.2] (p1) {\Ladiesroom};
    \node[ctrlPanel, right=0.2 of p1, label={[name=ctrl1l]}] (ctrl1) {};
    \node[big site] at ($($(p1)!0.5!(ctrl1)$) + (0,-0.4)$) (s1) {};
    \node[room, label={[name=r1l]}, fit=(ctrl1)(ctrl1l)(p1)(s1)] (r1) {};
    \node[big region, fit=(r1)(r1l)] (reg1) {};

    \node[big site, right=0.7 of ctrl1] (s2) {};
    \node[room, label={[name=r2l]}, fit=(s2)] (r2) {};
    \node[big region, fit=(r2)(r2l)] (reg2) {};

    \node[above right=0.1 of r1, yshift=1] (x) {\tiny $x$};
    \draw[big edge] (p1.east) -- (ctrl1.west);
    \draw[big edge] (ctrl1.east) to[out=0,in=-90] (x);
  \end{scope}
  \begin{scope}[local bounding box=rhs, shift={(2.8,0)}]
    \node[inner sep=0.2, opacity=0] (p1) {\Ladiesroom};
    \node[ctrlPanel, right=0.2 of p1, label={[name=ctrl1l]}] (ctrl1) {};
    \node[big site] at ($($(p1)!0.5!(ctrl1)$) + (0,-0.4)$) (s1) {};
    \node[room, label={[name=r1l]}, fit=(ctrl1)(ctrl1l)(p1)(s1)] (r1) {};
    \node[big region, fit=(r1)(r1l)] (reg1) {};

    \node[big site, right=0.7 of ctrl1] (s2) {};
    \node[inner sep=0.2, right=0.2 of s2] (p2) {\Ladiesroom};
    \node[room, label={[name=r2l]}, fit=(s2)(p2)] (r2) {};
    \node[big region, fit=(r2)(r2l)] (reg2) {};

    \node[above right=0.1 of r1, yshift=1] (x) {\tiny $x$};
    \draw[big edgec] (ctrl1.west) -- ($(ctrl1.west) + (-0.1,0)$);
    \draw[big edgec] (p2.east) -- ($(p2.east) + (0.1,0)$);
    \draw[big edge] (ctrl1.east) to[out=0,in=-90] (x);
  \end{scope}
  \node[] at ($($(lhs)!0.5!(rhs)$) + (-0.1,0)$) (rule) {$\rrul$};
\end{tikzpicture}
     }
    \caption{}
    \label{fig:rr_simple_rul}
  \end{subfigure}
  
	\begin{subfigure}[b]{\linewidth}
		\centering
    \begin{bigrapher}
atomic ctrl Person = 1;
atomic ctrl CtrlPanel = 2;
ctrl Room = 0;
ctrl Floor = 0;

react leave_secure =
 /z Room.(Person{z} | CtrlPanel{z,x} | id)
    || Room.id
 -->
    Room.(/z CtrlPanel{z,x} | id)
 || Room.(/z Person{z} | id);

big initialBigraph =
  /z Floor.(
      Room.(Person{z} | /x (Person{x} | /y CtrlPanel{x,y}))
    | Room.Person{z}
  );

begin brs
  init initialBigraph;
  rules = [ {leave_secure} ];
end
    \end{bigrapher}
		\caption{}
    \label{fig:rr_simple_big}
	\end{subfigure}
  \caption{BRS with a single reaction rule \rr{leave\_secure}.
  (a) Graphical notation for 
 \rr{leave\_secure}. (b) BigraphER snippet. }
  \label{fig:rr_simple}
\end{figure}

\begin{figure}
  \centering
   \resizebox{\linewidth}{!}{
     \begin{tikzpicture} [
  room/.append style = {draw},
  person/.append style = {inner sep=0.2},
  floor/.append style = {draw},
  ctrlPanel/.append style = {draw} ]
  \begin{scope}[local bounding box=lhs, shift={(0,0)}]
    \node[person, label={[inner sep=0.5, name=n2l]north:{}}] (n2) {\Ladiesroom};
    \node[person, right=0.8 of n2, label={[inner sep=0.5, name=n3l]north:{}}] (n3) {\Ladiesroom};
    \node[ctrlPanel, right=0.8 of n3, label={[inner sep=0.5, name=n4l]north:{}}] (n4) {};
    \node[person, right=0.8 of n4, label={[inner sep=0.5, name=n6l]north:{}}] (n6) {\Ladiesroom};
    \node[room, fit=(n4)(n4l)(n3)(n3l)(n2)(n2l), label={[inner sep=0.5, name=n1l]north:{\sf\tiny Room}}] (n1) {};
    \node[room, fit=(n6)(n6l), label={[inner sep=0.5, name=n5l]north:{\sf\tiny Room}}] (n5) {};
    \node[floor, fit=(n5)(n5l)(n1)(n1l), label={[inner sep=0.5, name=n0l]north:{\sf\tiny Floor}}] (n0) {};
    \node[big region, fit=(n0)(n0l)] (r0) {};
    \draw[big edge] (n2) to[out=0,in=-90, looseness=0.5] (n6);
    \draw[big edge] (n3) to[out=0,in=180] (n4);
    \draw[big edgec] (n4) to[out=0,in=180] ($(n4) + (0.3,0)$);
  \end{scope}
  \begin{scope}[local bounding box=rhs, shift={(5,0)}]
    \node[ctrlPanel, label={[inner sep=0.5, name=n2l]north:{\sf\tiny CtrlPanel}}] (n2) {};
    \node[person, right=0.8 of n2, label={[inner sep=0.5, name=n5l]north:{}}] (n5) {\Ladiesroom};
    \node[person, right=0.8 of n5, label={[inner sep=0.5, name=n4l]north:{}}] (n4) {\Ladiesroom};
    \node[person, right=0.8 of n4, label={[inner sep=0.5, name=n6l]north:{}}] (n6) {\Ladiesroom};
    \node[room, fit=(n5)(n5l)(n2)(n2l), label={[inner sep=0.5, name=n1l]north:{\sf\tiny Room}}] (n1) {};
    \node[room, fit=(n6)(n6l)(n4)(n4l), label={[inner sep=0.5, name=n3l]north:{\sf\tiny Room}}] (n3) {};
    \node[floor, fit=(n3)(n3l)(n1)(n1l), label={[inner sep=0.5, name=n0l]north:{\sf\tiny Floor}}] (n0) {};
    \node[big region, fit=(n0)(n0l)] (r0) {};
    \draw[big edgec] (n2) to[out=0,in=180] ($(n2) + (0.3,0)$);
    \draw[big edgec] (n2) to[out=180,in=0] ($(n2) + (-0.3,0)$);
    \draw[big edgec] (n4) to[out=0,in=-90] ($(n4) + (0.3,0.2)$);
    \draw[big edge] (n5) to[out=0,in=-90, looseness=0.5] (n6);
  \end{scope}

  \node[] at ($($(lhs)!0.5!(rhs)$) + (-0.1,0)$) (rule) {$\react$};
\end{tikzpicture}
   }
  \caption{Applying \rr{leave\_secure} to a simple scenario.}
  \label{fig:rr_leave_secure_app}
\end{figure}
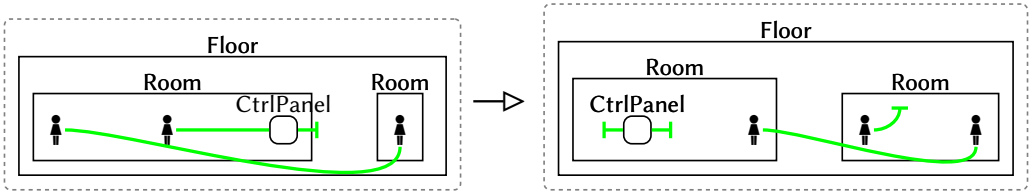

An example reaction rule, \rr{leave\_secure}, is in \cref{fig:rr_simple_rul}, with BigraphER representation in \cref{fig:rr_simple_big}.
In BigraphER we write rules with \biginline{-->} separating $L$ and $R$.
The rule describes how a \biginline{Person}, who is connected to a security \biginline{CtrlPanel}, can move between rooms.
Importantly, when moving out of the room containing the \biginline{CtrlPanel}, they sever their link to the \biginline{CtrlPanel} to avoid information leaks.
The result of applying \rr{leave\_secure} to a given bigraph is in \cref{fig:rr_leave_secure_app}.
Note there is only one valid occurrence for \rr{leave\_secure},  and so all other entities in the system are unchanged during the rewrite.

Sites/regions/names are especially important for rewriting as they allow a single rule to be applied in many circumstances.
In this case, the two sites allow any other bigraph to exist (including the empty bigraph) within the rooms, and intuitively we can think of the rule as saying ``find a room with \emph{at least} one \biginline{Person} and one \biginline{CtrlPanel} who are linked, and another \biginline{Room} that contains anything, including nothing''.
The use of parallel product (two \emph{regions}) means the two \biginline{Room}s cannot be a descendant of the other and do not need to be siblings (but can be), \ie the \biginline{Person} could move to a \biginline{Room} on a different floor, or even a different building.
The name $x$ allows the \biginline{CtrlPanel} to (possibly) be connected elsewhere, \eg to cameras in the building. If the link was closed (unnamed) the link must be an exact match.

Given a reaction rule, we define a BRS in BigraphER using a \biginline{begin brs ... end} block.
Inside this block we specify an initial bigraph (initial state), using the syntax \biginline{init b} (where \texttt{b} is a named bigraph), and set of rules as shown in \cref{fig:rr_simple_big}.
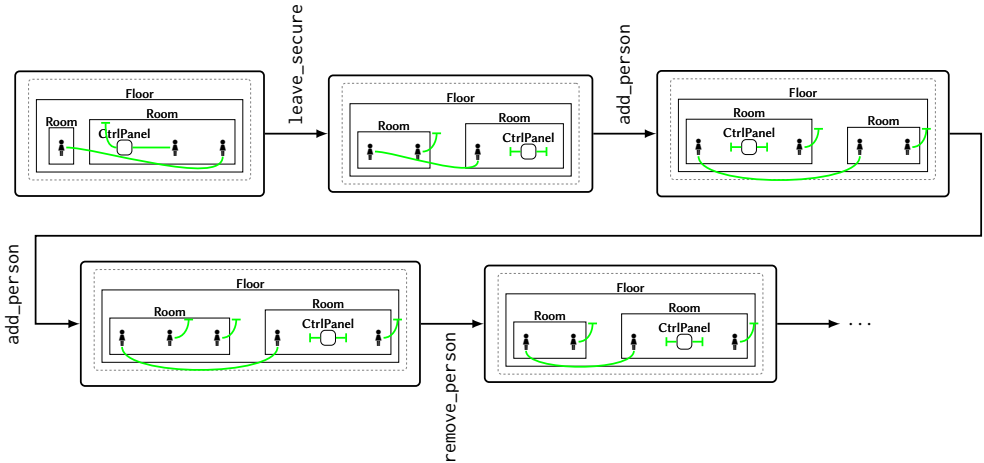
\begin{figure}
  \centering
   \resizebox{\linewidth}{!}{
     \newsavebox\szero \newsavebox\sone \newsavebox\stwo \newsavebox\sthree \newsavebox\sfour 

\sbox{\szero}{
\begin{tikzpicture}[
  ,
room/.append style = {draw},
person/.append style = {inner sep=0.2},
floor/.append style = {draw},
ctrlPanel/.append style = {draw}
  ]
    
\node[person,  label={[inner sep=0.5, name=n6l]north:{}}] (n6) {\Ladiesroom};
\node[ctrlPanel, right=0.80 of n6, label={[inner sep=0.5, name=n4l]north:{\sf\tiny CtrlPanel}}] (n4) {};
\node[person, right=0.60 of n4, label={[inner sep=0.5, name=n3l]north:{}}] (n3) {\Ladiesroom};
\node[person, right=0.60 of n3, label={[inner sep=0.5, name=n2l]north:{}}] (n2) {\Ladiesroom};
\node[room, fit=(n6)(n6l), label={[inner sep=0.5, name=n5l]north:{\sf\tiny Room}}] (n5) {};
\node[room, fit=(n4)(n4l)(n3)(n3l)(n2)(n2l), label={[inner sep=0.5, name=n1l]north:{\sf\tiny Room}}] (n1) {};
\node[floor, fit=(n5)(n5l)(n1)(n1l), label={[inner sep=0.5, name=n0l]north:{\sf\tiny Floor}}] (n0) {};
\node[big region, fit=(n0)(n0l)] (r0) {};
\draw[big edge] (n6) to[out=0,in=-90, looseness=0.5] (n2);
\draw[big edge] (n4) to[out=0,in=180] (n3);
\draw[big edgec] (n4) to[out=180,in=-90] ($(n4) + (-0.3,0.4)$);

\end{tikzpicture}
  
 }
\sbox{\sone}{
\begin{tikzpicture}[
  ,
room/.append style = {draw},
person/.append style = {inner sep=0.2},
floor/.append style = {draw},
ctrlPanel/.append style = {draw}
  ]
    
\node[person,  label={[inner sep=0.5, name=n6l]north:{}}] (n6) {\Ladiesroom};
\node[person, right=0.60 of n6, label={[inner sep=0.5, name=n4l]north:{}}] (n4) {\Ladiesroom};
\node[person, right=0.80 of n4, label={[inner sep=0.5, name=n5l]north:{}}] (n5) {\Ladiesroom};
\node[ctrlPanel, right=0.60 of n5, label={[inner sep=0.5, name=n2l]north:{\sf\tiny CtrlPanel}}] (n2) {};
\node[room, fit=(n6)(n6l)(n4)(n4l), label={[inner sep=0.5, name=n3l]north:{\sf\tiny Room}}] (n3) {};
\node[room, fit=(n5)(n5l)(n2)(n2l), label={[inner sep=0.5, name=n1l]north:{\sf\tiny Room}}] (n1) {};
\node[floor, fit=(n3)(n3l)(n1)(n1l), label={[inner sep=0.5, name=n0l]north:{\sf\tiny Floor}}] (n0) {};
\node[big region, fit=(n0)(n0l)] (r0) {};
\draw[big edgec] (n2) to[out=180,in=0] ($(n2) + (-0.3,0)$);
\draw[big edgec] (n2) to[out=0,in=180] ($(n2) + (0.3,0)$);
\draw[big edgec] (n4) to[out=0,in=-90] ($(n4) + (0.3,0.3)$);
\draw[big edge] (n6) to[out=0,in=-90, looseness=0.5] (n5);

\end{tikzpicture}
  
 }
\sbox{\stwo}{
\begin{tikzpicture}[
  ,
room/.append style = {draw},
person/.append style = {inner sep=0.2},
floor/.append style = {draw},
ctrlPanel/.append style = {draw}
  ]
    
\node[person,  label={[inner sep=0.5, name=n7l]north:{}}] (n7) {\Ladiesroom};
\node[ctrlPanel, right=0.60 of n7, label={[inner sep=0.5, name=n6l]north:{\sf\tiny CtrlPanel}}] (n6) {};
\node[person, right=0.60 of n6, label={[inner sep=0.5, name=n5l]north:{}}] (n5) {\Ladiesroom};
\node[person, right=0.80 of n5, label={[inner sep=0.5, name=n3l]north:{}}] (n3) {\Ladiesroom};
\node[person, right=0.60 of n3, label={[inner sep=0.5, name=n2l]north:{}}] (n2) {\Ladiesroom};
\node[room, fit=(n7)(n7l)(n6)(n6l)(n5)(n5l), label={[inner sep=0.5, name=n4l]north:{\sf\tiny Room}}] (n4) {};
\node[room, fit=(n3)(n3l)(n2)(n2l), label={[inner sep=0.5, name=n1l]north:{\sf\tiny Room}}] (n1) {};
\node[floor, fit=(n4)(n4l)(n1)(n1l), label={[inner sep=0.5, name=n0l]north:{\sf\tiny Floor}}] (n0) {};
\node[big region, fit=(n0)(n0l)] (r0) {};
\draw[big edgec] (n2) to[out=0,in=-90] ($(n2) + (0.3,0.3)$);
\draw[big edge] (n7) to[out=-90,in=-90, looseness=0.5] (n3);
\draw[big edgec] (n5) to[out=0,in=-90] ($(n5) + (0.3,0.3)$);
\draw[big edgec] (n6) to[out=0,in=180] ($(n6) + (0.3,0)$);
\draw[big edgec] (n6) to[out=180,in=0] ($(n6) + (-0.3,0)$);

\end{tikzpicture}
  
 }
\sbox{\sthree}{
\begin{tikzpicture}[
  ,
room/.append style = {draw},
person/.append style = {inner sep=0.2},
floor/.append style = {draw},
ctrlPanel/.append style = {draw}
  ]
    
\node[person,  label={[inner sep=0.5, name=n8l]north:{}}] (n8) {\Ladiesroom};
\node[person, right=0.60 of n8, label={[inner sep=0.5, name=n7l]north:{}}] (n7) {\Ladiesroom};
\node[person, right=0.60 of n7, label={[inner sep=0.5, name=n6l]north:{}}] (n6) {\Ladiesroom};
\node[person, right=0.80 of n6, label={[inner sep=0.5, name=n4l]north:{}}] (n4) {\Ladiesroom};
\node[ctrlPanel, right=0.60 of n4, label={[inner sep=0.5, name=n3l]north:{\sf\tiny CtrlPanel}}] (n3) {};
\node[person, right=0.60 of n3, label={[inner sep=0.5, name=n2l]north:{}}] (n2) {\Ladiesroom};
\node[room, fit=(n8)(n8l)(n7)(n7l)(n6)(n6l), label={[inner sep=0.5, name=n5l]north:{\sf\tiny Room}}] (n5) {};
\node[room, fit=(n4)(n4l)(n3)(n3l)(n2)(n2l), label={[inner sep=0.5, name=n1l]north:{\sf\tiny Room}}] (n1) {};
\node[floor, fit=(n5)(n5l)(n1)(n1l), label={[inner sep=0.5, name=n0l]north:{\sf\tiny Floor}}] (n0) {};
\node[big region, fit=(n0)(n0l)] (r0) {};
\draw[big edgec] (n2) to[out=0,in=-90] ($(n2) + (0.3,0.3)$);
\draw[big edgec] (n3) to[out=0,in=180] ($(n3) + (0.3,0)$);
\draw[big edgec] (n3) to[out=180,in=0] ($(n3) + (-0.3,0)$);
\draw[big edge] (n8) to[out=-90,in=-90, looseness=0.5] (n4);
\draw[big edgec] (n6) to[out=0,in=-90] ($(n6) + (0.3,0.3)$);
\draw[big edgec] (n7) to[out=0,in=-90] ($(n7) + (0.3,0.3)$);

\end{tikzpicture}
  
 }
\sbox{\sfour}{
\begin{tikzpicture}[
  ,
room/.append style = {draw},
person/.append style = {inner sep=0.2},
floor/.append style = {draw},
ctrlPanel/.append style = {draw}
  ]
    
\node[person,  label={[inner sep=0.5, name=n7l]north:{}}] (n7) {\Ladiesroom};
\node[person, right=0.60 of n7, label={[inner sep=0.5, name=n6l]north:{}}] (n6) {\Ladiesroom};
\node[person, right=0.80 of n6, label={[inner sep=0.5, name=n4l]north:{}}] (n4) {\Ladiesroom};
\node[ctrlPanel, right=0.60 of n4, label={[inner sep=0.5, name=n3l]north:{\sf\tiny CtrlPanel}}] (n3) {};
\node[person, right=0.60 of n3, label={[inner sep=0.5, name=n2l]north:{}}] (n2) {\Ladiesroom};
\node[room, fit=(n7)(n7l)(n6)(n6l), label={[inner sep=0.5, name=n5l]north:{\sf\tiny Room}}] (n5) {};
\node[room, fit=(n4)(n4l)(n3)(n3l)(n2)(n2l), label={[inner sep=0.5, name=n1l]north:{\sf\tiny Room}}] (n1) {};
\node[floor, fit=(n5)(n5l)(n1)(n1l), label={[inner sep=0.5, name=n0l]north:{\sf\tiny Floor}}] (n0) {};
\node[big region, fit=(n0)(n0l)] (r0) {};
\draw[big edgec] (n2) to[out=0,in=-90] ($(n2) + (0.3,0.3)$);
\draw[big edgec] (n3) to[out=0,in=180] ($(n3) + (0.3,0)$);
\draw[big edgec] (n3) to[out=180,in=0] ($(n3) + (-0.3,0)$);
\draw[big edge] (n7) to[out=-90,in=-90, looseness=0.5] (n4);
\draw[big edgec] (n6) to[out=0,in=-90] ($(n6) + (0.3,0.3)$);

\end{tikzpicture}
  
 }

\newcommand{\add}{\small\texttt{add\_person}\xspace}
\newcommand{\rem}{\small\texttt{remove\_person}\xspace}
\newcommand{\connect}{\small\texttt{connect}\xspace}
\newcommand{\leave}{\small\texttt{leave\_secure}\xspace}

\begin{tikzpicture}[
   state/.append style = { draw, rounded corners=2, thick, align=center },
   init/.append style = {}
  ]

  \node[state, init] (s0) {\usebox{\szero}};
  \node[state, right=1 of s0] (s1) {\usebox{\sone}};
  \node[state, right=1 of s1] (s2) {\usebox{\stwo}};
  \node[state, below=1 of s0, xshift=50] (s3) {\usebox{\sthree}};
  \node[state, right=1 of s3] (s4) {\usebox{\sfour}};
  \node[right=1 of s4] (more) {$\dots$};

  \draw[-latex, thick] (s0) -- node[right, rotate=90] {\leave} (s1);
  \draw[-latex, thick] (s1) -- node[right, rotate=90] {\add} (s2);

\draw[-latex, thick] (s2.east) -| ++(0.5,-1.6) -| node[left, midway, rotate=90, yshift=10] {\add} ([xshift=-20]s3.west) -- (s3.west);

  \draw[-latex, thick] (s3) -- node[left, rotate=90] {\rem} (s4);
  \draw[-latex, thick] (s4) -- (more);

\end{tikzpicture}
   }
  \caption{Applying a sequence of rewrite rules to an initial state (top left).}
  \label{fig:rr_trace}
\end{figure}

To show execution of a model, we give a possible sequence of rewrites in \cref{fig:rr_trace} for a BRS with the rule \rr{leave\_secure} and two additional rules \rr{add\_person} and \rr{remove\_person} allowing people to enter or exit the system so long as they are not connected elsewhere.
We use the same initial bigraph as in \cref{fig:rr_simple}.
This is one of many possible traces, \eg we could also apply \rr{add\_person} in the first state.
Notice that, as we use diagrammatic elements, we can move them around as we draw states so long as their relationships (nesting/linking) are maintained.

In practice, 
we require the left hand side of a rewrite  to be  
{\em solid}~\cite{DBLP:journals/entcs/KrivineMT08}\footnote{This ensures unique occurrences, which      is central to probabilistic and stochastic rewriting (\cref{sec:BRS2}).}.  

A bigraph is {\em solid} if
\begin{itemize}
\item  all regions contain at least one node and no outer names are idle 
\item no two sites or inner names are siblings 
\item   no site has a region as a parent
\item  no outer name is linked to an inner name.
\end{itemize}

These constraints only apply to the left hand side.  For example,   we can disconnect a name, \eg \biginline{A\{x\} --> /y A\{y\} | \{x\}}, resulting in an idle name on the right hand side ({\tt \{x\}}). BigraphER automatically enforces these constraints and we assume them throughout this paper.

 \subsection{Sites as Variables: Manipulating Sites During Rewriting}
\label{sec:instantiation}

\begin{figure}
    \centering
    \begin{subfigure}{0.49\linewidth}
    \centering
    \resizebox{0.8\linewidth}{!}{
      \begin{tikzpicture}[
    server/.append style = {draw},
    data/.append style = {draw, circle, fill=black, inner sep=0.5}
]
\begin{scope}[local bounding box=lhs, shift={(0,0)}]

  \node[data,  label={[inner sep=0.5, name=n1l]north:{\sf\tiny Data}}] (n1) {};
  \node[server, fit=(n1)(n1l), label={[inner sep=0.5, name=n0l]north:{\sf\tiny Server}}] (n0) {};
  \node[big region, fit=(n0)(n0l)] (r0) {};

\end{scope}
\begin{scope}[local bounding box=rhs, shift={($(lhs.east) + (1,-0.1)$)}]

  \node[server,  label={[inner sep=0.5, name=n0l]north:{\sf\tiny Server}}] (n0) {};
  \node[big region, fit=(n0)(n0l)] (r0) {};

\end{scope}

\node[xshift=0] at ($(lhs.east)!0.5!(rhs.west)$) {$\rrul$};

\begin{scope}[local bounding box=lhsb, shift={(-0.5,-1)}]

  \node[data,  label={[inner sep=0.5, name=n2l]north:{\sf\tiny Data}}] (n2) {};
  \node[data, right=0.60 of n2, label={[inner sep=0.5, name=n1l]north:{\sf\tiny Data}}] (n1) {};
  \node[server, fit=(n2)(n2l)(n1)(n1l), label={[inner sep=0.5, name=n0l]north:{\sf\tiny Server}}] (n0) {};
  \node[big region, fit=(n0)(n0l)] (r0) {};

\end{scope}
\begin{scope}[local bounding box=rhsb, shift={($(lhs.east) + (1.2,-1.1)$)}]

  \node[server,  label={[inner sep=0.5, name=n0l]north:{\sf\tiny Server}}] (n0) {};
  \node[big region, fit=(n0)(n0l)] (r0) {};

\end{scope}

\node[xshift=0] at ($(lhsb.east)!0.5!(rhsb.west)$) {$\rrul$};

\end{tikzpicture}
     }
    \caption{}
    \label{fig:inst_map_motivation_old}
    \end{subfigure}
    \begin{subfigure}{0.49\linewidth}
    \centering
    \resizebox{0.6\linewidth}{!}{
      \begin{tikzpicture}[
  ,
  server/.append style = {draw}
  ]
  \begin{scope}[local bounding box=lhs, shift={(0,0)}]

    \node[big site, ] (s0){};
    \node[server, fit=(s0), label={[inner sep=0.5, name=n0l]north:{\sf\tiny Server}}] (n0) {};
    \node[big region, fit=(n0)(n0l)] (r0) {};

  \end{scope}
  \begin{scope}[local bounding box=rhs, shift={($(lhs.east) + (1,-0.1)$)}]

    \node[server,  label={[inner sep=0.5, name=n0l]north:{\sf\tiny Server}}] (n0) {};
    \node[big region, fit=(n0)(n0l)] (r0) {};

  \end{scope}

  \node[xshift=0] at ($(lhs.east)!0.5!(rhs.west)$) {$\rrul$};
\end{tikzpicture}
     }
    \caption{}
    \label{fig:inst_map_motivation_site}
    \end{subfigure}
    \caption{(a) Delete requiring multiple rules.  (b) Delete as a single rule using a site.}
    \label{fig:inst_map_motivation}
\end{figure}
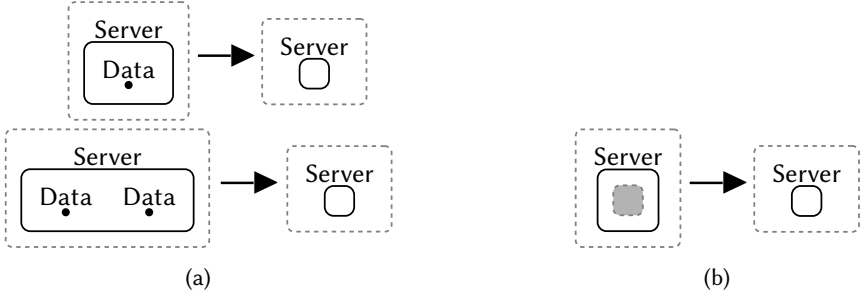

Sites are a  powerful feature of bigraphs because they act like variables. 
For example, consider a rule to delete data from a server as shown in \cref{fig:inst_map_motivation}.
Without sites, we are forced to include rules to delete each number of data items as in \cref{fig:inst_map_motivation_old}.

Sites allow us to specify  a set  of bigraphs without explicitly enumerating   all the elements.
That is, one rule can apply in many situations.
For example,  we can delete any number of data items (and anything else on the server) using a single rule shown in \cref{fig:inst_map_motivation_site}.
In this case, we have \emph{deleted} a site and the entire bigraph it abstracted over is removed.

\tipbox{Sites are abstractions over bigraphs, i.e. they are bigraph variables that can be instantiated with a bigraph.  Sites should be used when defining general rules that apply in many situations.}

During rewriting we can \emph{duplicate} or \emph{discard} the contents of a site using a special construct known as an \emph{instantiation map}.
These operations occur frequently in practical bigraph models.

We identify sites numerically based on their position in a rule definition, with the left-most site being site 0.
For example in \biginline{A.id || B.id} the site below \biginline{A} is site 0 and below \biginline{B} is site 1.
An instantiation map determines, for each site in the right-hand-side of a rule, which sites this maps to on the left-hand-side of a rule.
For example, an instantiation map: $0 \mapsto 1$, $1 \mapsto 0$, gives site $0$ on the right, the contents of site $1$ on the left, and site $1$ on the right, the contents of site $0$ on the left (\ie it implements a swap).
All sites on the right-hand-side must correspond to a site on the left, but not all left-hand sites need to be included.
For example, the map $0 \mapsto 0$, $1 \mapsto 0$, would duplicate site $0$ from the left into both right-hand sites and the remaining site (site 1) on the left is discarded.

We write instantiation maps at the end of a rule definition using the syntax \biginline{@[n,...,m]} (for natural numbers \biginline{n, m}).
The value of the $i^{th}$ element of this list determines the left-hand site that the $i^{th}$ right-hand site corresponds to.
For example \biginline{@[0,1,1]} maps sites $0 \mapsto 0$, $1 \mapsto 1$, $2 \mapsto 1$.
This map must be fully defined and so the length should match the number of sites on the right hand side.
This avoids the situation where we have a site but no information about how to instantiate it. 

Graphically, we draw instantiation maps using  blue dashed arrows.
For clarity, we may sometimes  draw only select arrows when it is a 1-to-1 mapping, other than specific sites, and it is clear from the rule what we intend.

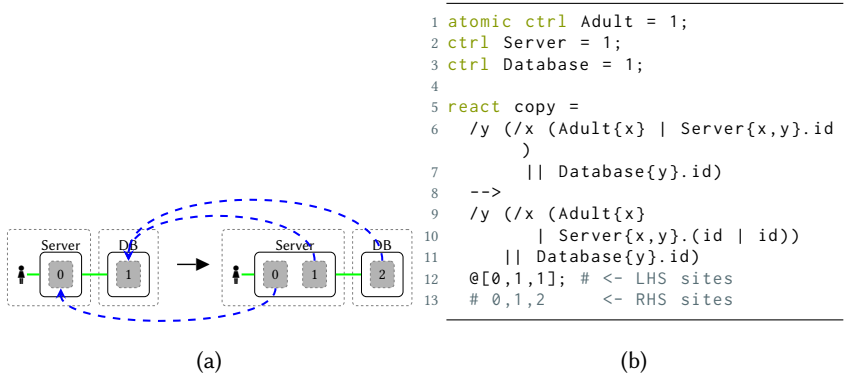
\begin{figure}
  \centering
    \begin{subfigure}[b]{0.4\linewidth}
        \centering
    \resizebox{\linewidth}{!}{
      \begin{tikzpicture}
  \begin{scope}[local bounding box=lhs, shift={(0,0)}]
    \node[inner sep=0.2] (p1) {\Ladiesroom};
    \node[big site, right=0.3 of p1] (s1l) {\tiny 0};
    \node[server, label={[inner sep=0.5, name=serv1l]north:{\sf\tiny Server}}, fit=(s1l)] (serv1) {};
    \node[big region, fit=(p1)(serv1)(serv1l)] (r1) {};

    \node[big site, right=0.6 of s1l] (s2l) {\tiny 1};
    \node[server, label={[inner sep=0.5, name=serv2l]north:{\sf\tiny DB}}, fit=(s2l)] (serv2) {};
    \node[big region, fit=(serv2)(serv2l)] (r2) {};

    \draw[big edge] (p1.east) to[out=0, in=180] (serv1.west);
    \draw[big edge] (serv1.east) to[out=0, in=180] (serv2.west);
  \end{scope}

  \begin{scope}[local bounding box=rhs, shift={(3,0)}]
    \node[inner sep=0.2] (p1) {\Ladiesroom};
    \node[big site, right=0.3 of p1] (s1r) {\tiny 0};
    \node[big site, right=0.2 of s1r] (s2r) {\tiny 1};
    \node[server, label={[inner sep=0.5, name=serv1l]north:{\sf\tiny Server}}, fit=(s1r)(s2r)] (serv1) {};
    \node[big region, fit=(p1)(serv1)(serv1l)] (r1) {};

    \node[big site, right=0.6 of s2r] (s3r) {\tiny 2};
    \node[server, label={[inner sep=0.5, name=serv2l]north:{\sf\tiny DB}}, fit=(s3r)] (serv2) {};
    \node[big region, fit=(serv2)(serv2l)] (r2) {};

    \draw[big edge] (p1.east) to[out=0, in=180] (serv1.west);
    \draw[big edge] (serv1.east) to[out=0, in=180] (serv2.west);
  \end{scope}

  \draw[big inst map] (s1r) to[out=-90, in=-90, looseness=0.5] (s1l);
  \draw[big inst map] (s2r) to[out=90, in=90, looseness=0.8] (s2l);
  \draw[big inst map] (s3r) to[out=90, in=90, looseness=0.8] (s2l);

  \node[] at ($($(lhs)!0.5!(rhs)$) + (-0.1,0)$) (rule) {$\rrul$};
\end{tikzpicture}
     }
        \caption{}
    \end{subfigure}
    \begin{subfigure}[b]{0.4\linewidth}
        \centering
    \begin{bigrapher}
atomic ctrl Adult = 1;
ctrl Server = 1;
ctrl Database = 1;

react copy =
  /y (/x (Adult{x} | Server{x,y}.id) 
       || Database{y}.id)
  -->
  /y (/x (Adult{x} 
        | Server{x,y}.(id | id))
     || Database{y}.id)
  @[0,1,1]; # <- LHS sites
  # 0,1,2     <- RHS sites
    \end{bigrapher}
        \caption{}
    \end{subfigure}
  \caption{Copy through instantiation map.}
  \label{fig:inst_map}
  \end{figure}
  
We show the power of instantiation maps by example, including how they let us model \emph{copy} and \emph{delete} operations.
Movement of entities happens often in physical scenarios, \eg moving between rooms, but copying and deletion of entities is less common. 
To show these features of instantiation maps we extend our building example with servers and data.

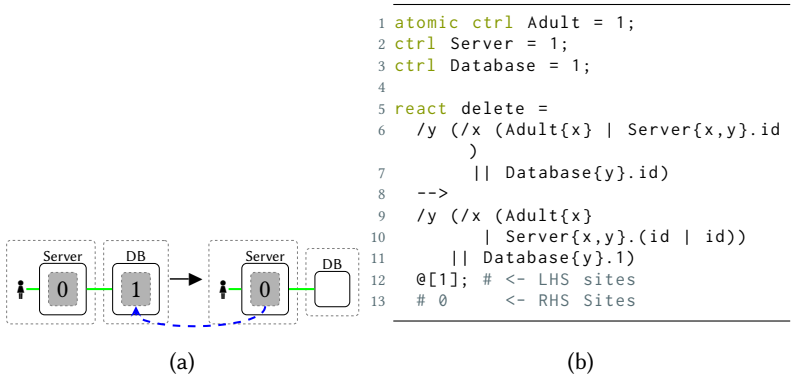
\begin{figure}
  \centering
    \begin{subfigure}[b]{0.35\linewidth}
        \centering
    \resizebox{\linewidth}{!}{
      \begin{tikzpicture}
  \begin{scope}[local bounding box=lhs, shift={(0,0)}]
    \node[inner sep=0.2] (p1) {\Ladiesroom};
    \node[big site, right=0.3 of p1] (s1l) {0};
    \node[server, label={[inner sep=0.5, name=serv1l]north:{\sf\tiny Server}}, fit=(s1l)] (serv1) {};
    \node[big region, fit=(p1)(serv1)(serv1l)] (r1) {};

    \node[big site, right=0.6 of s1l] (s2l) {1};
    \node[server, label={[inner sep=0.5, name=serv2l]north:{\sf\tiny DB}}, fit=(s2l)] (serv2) {};
    \node[big region, fit=(serv2)(serv2l)] (r2) {};

    \draw[big edge] (p1.east) to[out=0, in=180] (serv1.west);
    \draw[big edge] (serv1.east) to[out=0, in=180] (serv2.west);
  \end{scope}

  \begin{scope}[local bounding box=rhs, shift={(2.8,0)}]
    \node[inner sep=0.2] (p1) {\Ladiesroom};
    \node[big site, right=0.3 of p1] (s1r) {0};
    \node[server, label={[inner sep=0.5, name=serv1l]north:{\sf\tiny Server}}, fit=(s1r)] (serv1) {};
    \node[big region, fit=(p1)(serv1)(serv1l)] (r1) {};

    \node[big site, right=0.6 of s1r, opacity=0] (s2r) {};
    \node[server, label={[inner sep=0.5, name=serv2l]north:{\sf\tiny DB}}, fit=(s2r)] (serv2) {};
    \node[big region, fit=(serv2)(serv2l)] (r2) {};

    \draw[big edge] (p1.east) to[out=0, in=180] (serv1.west);
    \draw[big edge] (serv1.east) to[out=0, in=180] (serv2.west);
  \end{scope}

  \draw[big inst map] (s1r) to[out=-90, in=-90, looseness=0.5] (s2l);

  \node[] at ($($(lhs)!0.5!(rhs)$) + (0,0)$) (rule) {$\rrul$};
\end{tikzpicture}
     }
        \caption{}
    \end{subfigure}
    \begin{subfigure}[b]{0.4\linewidth}
        \centering
    \begin{bigrapher}
atomic ctrl Adult = 1;
ctrl Server = 1;
ctrl Database = 1;

react delete =
  /y (/x (Adult{x} | Server{x,y}.id) 
       || Database{y}.id)
  -->
  /y (/x (Adult{x} 
        | Server{x,y}.(id | id))
     || Database{y}.1)
  @[1]; # <- LHS sites
  # 0     <- RHS Sites
    \end{bigrapher}
        \caption{}
    \end{subfigure}
  \caption{Copy-and-Delete through an instantiation map.}
  \label{fig:rr_inst_delete}
\end{figure}

An example rule using an instantiation map is in \cref{fig:inst_map}.
In this case we \emph{copy} everything that was in the database (including nothing if it was empty) to the local server, while keeping all local server data intact.
To make this mapping clear we have numbered the sites in this example, but will not number sites in general.
Similarily, by just changing the instantiation map, we model copy-and-delete in \cref{fig:rr_inst_delete}.
In this case the site under the \biginline{Database} is not in the map and so is dropped.

\tipbox{Careful consideration needs to be given to duplicating sites when the bigraphs being
duplicated contain links: when   a site is duplicated that contains links,  the links remain connected. For example, if we copy \biginline{A\{x\}} to obtain \biginline{A\{x\} | A\{x\}}, then both  \biginline{A} entities are connected in the result\footnote{This is not the case for binding bigraphs~\cite[11.3]{milner_SpaceAndMotionOfCommunicatingAgents:2009} where a name can be specified as local to some entity.}.}

\section{Parameterised, instantaneous, and conditional    rules }\label{sec:extension1}
We have extended bigraph entities and  rewriting in several ways, all of which have been implemented in  BigraphER.

\subsection{Parameterised Entities}
\label{sec:parameterised}
 Models may require numeric operations or to  assign identifiers to entities, \eg
\biginline{Person(1)},
\biginline{Person(2)}, \etc.
While it is possible to encode numbers using schemes such as Peano arithmetic, these can be difficult to work with and have significant computational overhead.

 BigraphER provides \emph{parameterised entities}, \eg \biginline{Nat(n)}, $n \in \mathbb{N}$ that represent \emph{families} of entities, one for each value of $n$. We also support float and string parameters. 
  We   view this as syntactic sugar for defining a set of entities: \biginline{Nat(0)}, \biginline{Nat(1)}, \dots.
For practical models we choose $n$ as finite (and defined by the user), although the theory supports infinite entity sets if required.
Entities can vary in multiple parameters if required, \eg \biginline{A(n,m)}.
We use the syntax \biginline{fun ctrl A(x) = 0} (where \biginline{x} is an arbitrary identifier) to denote a parameterised entity\footnote{\biginline{fun} is a reference to function, \ie it takes parameters and produces new entities.}.

\subsection{Parameterised Rules}
Parameterised rules represent a \emph{family} of rules.
For example, we can write \rrP{r}{n} for $n \in \mathbb{N}$ in place of \rrP{r}{0}, \rrP{r}{1}, \dots.
Within the rules we allow the bound variables to be used for parameterised entities.

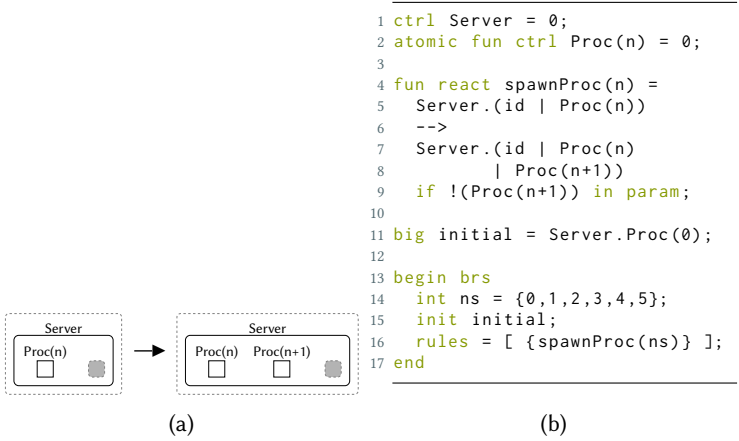
\begin{figure}
  \centering
	\begin{subfigure}[b]{0.35\linewidth}
		\centering
    \resizebox{\linewidth}{!}{
      \begin{tikzpicture}[
  server/.append style = {draw},
  process/.append style = {draw} ]
  \begin{scope}[local bounding box=lhs, shift={(0,0)}]
    \node[process,  label={[inner sep=0.5, name=n1l]north:{\sf\tiny Proc(n)}}] (n1) {};
    \node[big site, right=0.500000 of n1,] (s0){};
    \node[server, fit=(n1)(n1l)(s0), label={[inner sep=0.5, name=n0l]north:{\sf\tiny Server}}] (n0) {};
    \node[big region, fit=(n0)(n0l)] (r0) {};
  \end{scope}
  \begin{scope}[local bounding box=rhs, shift={(2.5,0)}]
    \node[process,  label={[inner sep=0.5, name=n1l]north:{\sf\tiny Proc(n)}}] (n1) {};
    \node[process, right=0.700000 of n1, label={[inner sep=0.5, name=n2l]north:{\sf\tiny Proc(n+1)}}] (n2) {};
    \node[big site, right=0.500000 of n2,] (s0){};
    \node[server, fit=(n2)(n2l)(n1)(n1l)(s0), label={[inner sep=0.5, name=n0l]north:{\sf\tiny Server}}] (n0) {};
    \node[big region, fit=(n0)(n0l)] (r0) {};
  \end{scope}

  \node[] at ($($(lhs.east)!0.5!(rhs.west)$) + (-0,0)$) (rule) {$\rrul$};
\end{tikzpicture}
     }
		\caption{}
	\end{subfigure}
	\begin{subfigure}[b]{0.35\linewidth}
		\centering
    \begin{bigrapher}
ctrl Server = 0;
atomic fun ctrl Proc(n) = 0;

fun react spawnProc(n) =
  Server.(id | Proc(n))
  -->
  Server.(id | Proc(n) 
         | Proc(n+1))
  if !(Proc(n+1)) in param;
  
big initial = Server.Proc(0);

begin brs
  int ns = {0,1,2,3,4,5};
  init initial;
  rules = [ {spawnProc(ns)} ];
end
    \end{bigrapher}
		\caption{}
	\end{subfigure}
  \caption{Example parameterised rule. (a) Rule for  spawning server processes. (b) BigraphER snippet.} 
  \label{fig:param_rule}
\end{figure}

As with parameterised entities, in practice parameterised rule variables must be instantiated with a finite set of values.
In BigraphER we allow numeric operations on parameters to be applied within a rule. These operations are performed as the model is compiled, \ie the resulting rules are fully instantiated.

A simple parameterised rule is in \cref{fig:param_rule} where we 
 allow processes running on a server to spawn future processes:  \biginline{Proc(n)} spawns 
 \biginline{Proc(n+1)}, up to $n=5$.   
 
In this case there is no rule \biginline{spawn_proc(6)}.
We use the syntax \biginline{fun react} to define a parameterised rule instead of a standard rule.

\tipbox{
Parameterised rules are syntactic sugar for a set of underlying rules, so   use them sparingly as each new rule increases the work needed for system analysis. In practice, this   affects   how you  describe   entities. For example, it is often better to define \biginline{Camera.CName(1)} 
which can be abstracted by a site whenever the identifier is unimportant, \eg \biginline{Camera.id},
rather than \biginline{Camera(1)},  which requires a family of rules \emph{every} time a rule uses a \biginline{Camera}.}  
We have extended bigraph rewriting in several ways, all of which have been implemented in  BigraphER.

\subsection{Rule priorities}
\label{sec:priorities}

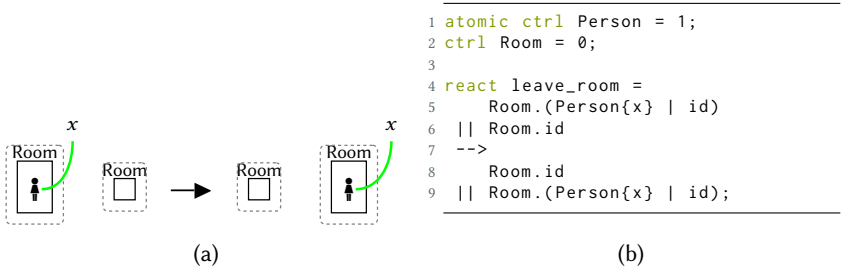
\begin{figure}
  \centering
\begin{subfigure}[b]{0.4\linewidth}
		\centering
    \resizebox{\linewidth}{!}{
      \begin{tikzpicture}[
room/.append style = {draw},
person/.append style = {inner sep=0.2} ]

 \begin{scope}[local bounding box=lhs, shift={(0,0)}]
   \node[person, label={[inner sep=0.5, name=n1l]north:{}}] (n1) {\Ladiesroom};
   \node[room, right=0.800000 of n1, label={[inner sep=0.5, name=n2l]north:{}}] (n2) {};
   \node[room, fit=(n1)(n1l), label={[inner sep=0.5, name=n0l]north:{}}] (n0) {};
   \node[big region, fit=(n0)(n0l)] (r0) {};
   \node[big region, fit=(n2)(n2l)] (r1) {};
   \node[yshift=20,xshift=12] (name_l) {\tiny $x$};
   \draw[big edge] (n1) to[out=0,in=-90] (name_l);

  \end{scope}
  \begin{scope}[local bounding box=rhs, shift={(2.5,0)}]
    \node[room, label={[inner sep=0.5, name=n0l]north:{}}] (n0) {};
    \node[person, right=0.8000000 of n0, label={[inner sep=0.5, name=n2l]north:{}}] (n2) {\Ladiesroom};
    \node[room, fit=(n2)(n2l), label={[inner sep=0.5, name=n1l]north:{}}] (n1) {};
    \node[big region, fit=(n0)(n0l)] (r0) {};
    \node[big region, fit=(n1)(n1l)] (r1) {};
    \node[yshift=20,xshift=42] (name_l) {\tiny $x$};
    \draw[big edge] (n2) to[out=0,in=-90] (name_l);
  \end{scope}

  \node[] at ($($(lhs)!0.5!(rhs)$) + (-0.1,-0.3)$) (rule) {$\rrul$};
\end{tikzpicture}
     }
    \caption{}
    \label{fig:rr_leave_room_fig}
  \end{subfigure}
	\begin{subfigure}[b]{0.4\linewidth}
		\centering
    \begin{bigrapher}
atomic ctrl Person = 1;
ctrl Room = 0;

react leave_room =
    Room.(Person{x} | id)
 || Room.id
 -->
    Room.id
 || Room.(Person{x} | id);
    \end{bigrapher}
		\caption{}
    \label{rr_leave_room_big}
	\end{subfigure}
  \caption{Leaving a room. (a) Reaction rule \rr{leave\_room}.
  (b) BigraphER snippet}
  \label{fig:rr_leave_room}
\end{figure}

In general a reaction rule can be applied whenever there is a suitable match.
However in practice we often want more control over when rules can be applied.
For example, consider the rule \rr{leave\_room} in \cref{fig:rr_leave_room_fig}.
This rule generalises \rr{leave\_secure} (\cref{fig:rr_simple_rul}) allowing people to move between between arbitrary rooms.
Due to the connection to name \biginline{x}, this means a person connected to a \biginline{CtrlPanel} might leave the room and still be connected to the security network: an information leak.
To avoid this, \rr{leave\_secure} can be applied with higher priority than \rr{leave\_room}.

A better way to model the above would be to have a single movement rule \rr{leave\_room} and a high priority rule that \emph{fixes} the model after a movement (before any other rule applies) by severing any links to control panels in other rooms.
This is shown in \cref{fig:rr_fix_leave_room}.
This way we do not block other entities from moving.

\begin{figure}
  \centering
	\begin{subfigure}[b]{\linewidth}
		\centering
    \resizebox{0.8\linewidth}{!}{
      \begin{tikzpicture} [
  room/.append style = {draw},
  person/.append style = {inner sep=0.2},
  ctrlPanel/.append style = {draw}
  ]
  \begin{scope}[local bounding box=lhs, shift={(0,0)}]

    \node[person,  label={[inner sep=0.5, name=n1l]north:{}}] (n1) {\Ladiesroom};
    \node[big site, right=0.3 of n1,] (s0){};
    \node[ctrlPanel, right=0.9 of s0, label={[inner sep=0.5, name=n3l]north:{\sf\tiny CtrlPanel}}] (n3) {};
    \node[big site, right=0.3 of n3,] (s1){};
    \node[room, fit=(n1)(n1l)(s0), label={[inner sep=0.5, name=n0l]north:{\sf\tiny Room}}] (n0) {};
    \node[room, fit=(n3)(n3l)(s1), label={[inner sep=0.5, name=n2l]north:{\sf\tiny Room}}] (n2) {};
    \node[big region, fit=(n0)(n0l)] (r0) {};
    \node[big region, fit=(n2)(n2l)] (r1) {};
    \node[yshift=25,xshift=23] (name_y) {\tiny $y$};
    \draw[big edge] (n1) to[out=-90,in=-90, looseness=0.5] (n3);
    \draw[big edge] (n3) to[out=180,in=-90] (name_y);
  \end{scope}

  \begin{scope}[local bounding box=rhs, shift={(4,0)}]
    \node[person,  label={[inner sep=0.5, name=n1l]north:{}}] (n1) {\Ladiesroom};
    \node[big site, right=0.3 of n1,] (s0){};
    \node[ctrlPanel, right=0.9 of s0, label={[inner sep=0.5, name=n3l]north:{\sf\tiny CtrlPanel}}] (n3) {};
    \node[big site, right=0.3 of n3,] (s1){};
    \node[room, fit=(n1)(n1l)(s0), label={[inner sep=0.5, name=n0l]north:{\sf\tiny Room}}] (n0) {};
    \node[room, fit=(n3)(n3l)(s1), label={[inner sep=0.5, name=n2l]north:{\sf\tiny Room}}] (n2) {};
    \node[big region, fit=(n0)(n0l)] (r0) {};
    \node[big region, fit=(n2)(n2l)] (r1) {};
    \node[yshift=25,xshift=23] (name_y) {\tiny $y$};
    \draw[big edgec] (n1) to[out=0,in=180] ($(n1) + (0.3,0)$);
    \draw[big edgec] (n3) to[out=0,in=180] ($(n3) + (0.3,0)$);
    \draw[big edge] (n3) to[out=180,in=-90] (name_y);
  \end{scope}

  \node[] at ($($(lhs)!0.5!(rhs)$) + (-0,-0.2)$) (rule) {$\rrul$};
\end{tikzpicture}
     }
		\caption{}
	\end{subfigure}
	
	\begin{subfigure}[b]{\linewidth}
		\centering
    \begin{bigrapher}
react fix_secure =
 /x (Room.(Person{x} | id)
 ||  Room.(CtrlPanel{x,y} | id))
    -->
    Room.(/x Person{x} | id)
 || Room.(/x CtrlPanel{x,y} | id);

begin brs
  init ...;
  rules = [ {fix_secure}, {leave_room} ];
end
    \end{bigrapher}
		\caption{}
	\end{subfigure}
  \caption{Using priorities to ensure links to control panels are severed without stopping movement. (a) Rule \biginline{leave\_room}.
  (b) BigraphER snippet.}
  \label{fig:rr_fix_leave_room}
\end{figure}

BigraphER   allows \emph{rule priorities} that define a partial ordering on rule application.
Each reaction rule belongs to one \emph{priority class} (sets of rules with the same priority) and an operator $<$ defines the partial order.
For example, we can define $\{X\} < \{Y\}$ meaning that rules in $X$ are checked for matches only if no rule in $Y$ has a match.
Within a priority class rules apply non-deterministically, as before. In BigraphER syntax the classes are enumerated in the rule declaration within \biginline{\{\ \}} brackets, \eg in 
\cref{fig:rr_fix_leave_room}b there are two classes, one with higher priority containing 
\biginline{fix\_secure}  and the lower priority class containing 
\biginline{leave\_room}. 

\tipbox{Be careful when assigning rule priorities. Although priorities stop a general case applying when a specific should be applied, it also stops a general case being applied to any \emph{other} matches.}
While it is possible to encode priorities directly in the bigraphs through additional entities, this: i)~causes larger models, 
ii)~mixes domain specific modelling entities and control-only entities leading to more complex models, and  iii)~is error prone.

\subsection{Instantaneous rules}
\label{sec:instantaneous}

BigraphER supports \emph{instantaneous rules} that allow a set of reactions to occur without adding intermediate states to the transition system.
The rules are specified in the standard manner and   placed in an instantaneous priority class, denoted in BigraphER with \biginline{(\ )} instead of \biginline{\{\ \}}.
Instantaneous priority classes are special in that they must \emph{fully reduce} before any additional rules are called, \ie $\{ r_{1}, r_{2}\}$ will apply either $r_{1}$ or $r_{2}$ (if possible) and then retry rules with higher priority; $(r_{1}, r_{2})$  applies \emph{continuously} $r_{1}$ or $r_{2}$ until no further applications are possible.
 Instantaneous rules  must be confluent,  that is, they must always create the same resulting bigraph regardless of  the order of   application. The result  is a more efficient transition system in which    many spurious states and interleavings are removed.

\subsection{Conditional rules}
\label{sec:conditionals}

We often   want   to control the \emph{contexts} under which a reaction rule can apply.
This is possible in Conditional Bigraphs~\cite{archibald.ea_ConditionalBigraphs:2020}, supported by BigraphER, that allow additional conditions---specified as matches---to be attached to a rule.

A match for a rewrite has three components: the left-hand-side of the rule, a \emph{parameter} that is everything \emph{inside} the sites, and a \emph{context} that is everything other than the rule and parameter.
Conditional rules lets us constrain either the parameter (most commonly) or the context.

\begin{figure}
  \centering
	\begin{subfigure}[b]{0.4\linewidth}
		\centering
    \resizebox{\linewidth}{!}{
      \begin{tikzpicture}[
  server/.append style = {draw},
  room/.append style = {draw},
  employee/.append style = {draw} ]
  \begin{scope}[local bounding box=lhs, shift={(0,0)}]
    \node[employee, label={[inner sep=0.5, name=n1l]north:{\sf\tiny Employee}}] (n1) {};
    \node[server, right=0.600000 of n1, label={[inner sep=0.5, name=n2l]north:{\sf\tiny Server}}] (n2) {};
    \node[big site, right=0.4 of n2] (s1) {};
    \node[room, fit=(n2)(n2l)(n1)(n1l)(s1), label={[inner sep=0.5, name=n0l]north:{\sf\tiny Room}}] (n0) {};
    \node[big region, fit=(n0)(n0l)] (r0) {};
    \draw[big edgec] (n1) to[out=0,in=180] ($(n1) + (0.3,0)$);
    \draw[big edgec] (n2) to[out=0,in=180] ($(n2) + (0.3,0)$);
  \end{scope}
  \begin{scope}[local bounding box=rhs, shift={(3.2,0)}]
    \node[employee, label={[inner sep=0.5, name=n1l]north:{\sf\tiny Employee}}] (n1) {};
    \node[server, right=0.600000 of n1, label={[inner sep=0.5, name=n2l]north:{\sf\tiny Server}}] (n2) {};
    \node[big site, right=0.4 of n2] (s1) {};
    \node[room, fit=(n2)(n2l)(n1)(n1l)(s1), label={[inner sep=0.5, name=n0l]north:{\sf\tiny Room}}] (n0) {};
    \node[big region, fit=(n0)(n0l)] (r0) {};
    \draw[big edge] (n1) to[out=0,in=180] (n2);
  \end{scope}

  \node[] at ($($(lhs)!0.5!(rhs)$) + (0,0)$) (rule) {$\rrul$};
\end{tikzpicture}
     }
    \caption{}
    \label{fig:rr_connect_server}
  \end{subfigure}
	\begin{subfigure}[b]{0.4\linewidth}
		\centering
    \begin{bigrapher}
atomic ctrl Employee = 1;
atomic ctrl Visitor = 0;
ctrl Room = 0;
ctrl Server = 1;

react connect_server =
  Room.( /x Employee{x} 
       | /s Server{s}.id 
       | id)
  -->
  Room.(/x
         ( Employee{x}  
         | Server{x}.id) 
       | id)
  if !Visitor in param;
    \end{bigrapher}
		\caption{}
    \label{fig:big_connect_server}
	\end{subfigure}
	\label{fig:conditional_example}
  \caption{Conditional rule \rr{connect\_server}.
  (a) connect\_server rule. (b) BigraphER snippet.}
\end{figure}
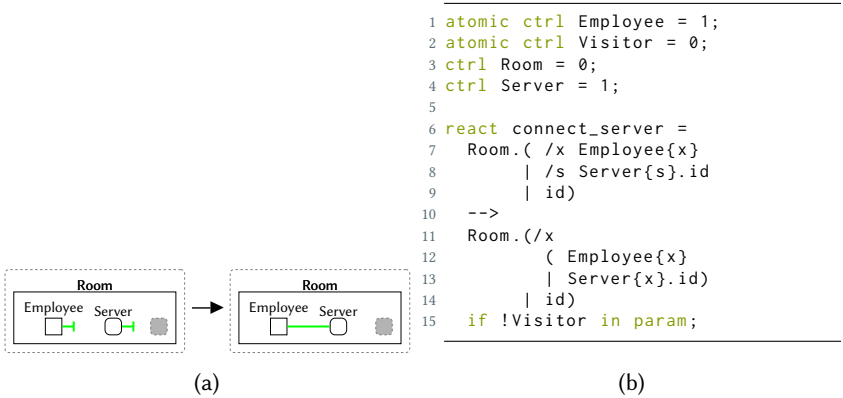

An example conditional rule is in \cref{fig:conditional_example}.
 
In this example we want to avoid a leak of sensitive information by disallowing employees access to a server if there are any visitors in the room.
We   specify this as a condition that states the rule only applies when there are no \biginline{Visitor} entities in the \emph{parameter}, \ie the site.

We use the notation \biginline{if !X in param} to specify that \biginline{X} should not appear in the parameter. 
Similarly we can have conditionals that \emph{require} an entity to be present, \eg \biginline{if Y in ctx} means there must be a \biginline{Y} somewhere in the model that is not in the rule or parameter.

\tipbox{Conditional rules allow us to restrict instantiation of sites: instead of allowing arbitrary bigraphs, conditions allow only those that do/do not match a pattern.}

There are three main caveats when using conditions: i)~conditions cannot determine a \emph{specific} site (or region) to match within, and need to be valid regardless of where the condition is located in the parameter, \ie we cannot say site 0 does not contain a \biginline{Person} only no site contains a \biginline{Person}; ii)~names in a conditional, even if they are the same as those in the match, are not guaranteed to be connected---this is due to names being structural elements (see \cref{sec:named}) rather than global, \ie we can rename in the conditional as required, and; iii)~conditions cannot be nested, \ie a condition cannot itself have a condition.

\tipbox{Where possible, it is better to choose a conditional rule over rule priorities. This is because conditions indicate the intent of a single rule, while priorities  define relationships across the whole set of rewrite rules.}

This  concludes our overview of    BRS, 
  the following three sections cover a range of practical modelling advice for    tackling common   scenarios.
Creating models is as much an art as a science, and these techniques should be seen as pieces of advice rather than absolute rules.

\section{Multi-perspective modelling}\label{multi}
\label{sec:multiperspective}

The use of parallel regions, with links that can cross regions,     provides a way to construct models with a strong separation of concerns.
We   can consider each parallel region to be a \emph{perspective}, for example we might have a region that models the \emph{physical} characteristics of a system, while another might model \emph{virtual} representations. 
Cross-region links between entities allow  them to be related, \eg to tie a \emph{physical} server to \emph{virtual} characteristics. Note that rewrite rules can be over multiple perspectives. 
As links are undirected hyperlinks, the relationships between perspectives form a \emph{graph} \ie we can have multiple representations of the same entity.

Multi-perspective modelling has been used to good effect in previous works.
For example,  the model of a cyber-physical game, Savannah~\cite{benford.ea_Savannah:2016}, is split into four {\em design} perspectives: \emph{physical}, that models people entering and exiting the physical game space (a playing field); \emph{human}, that models  proxemics and how  human players interact to form groups; \emph{technology}, that models how GPS technology senses and represents the  physical reality,   including drift and ghosting; and \emph{computational}, that  implements the rules of the game. The model was used to study behaviours observed in user trials, in particular player  cognitive dissonance   in certain situations. 
Another example is a model for a sensor system that is split into three perspectives: \emph{physical}, where each sensor is located; \emph{control}, information on sensor capabilities; and \emph{application}, the devices an application requires to operate~\cite{sevegnani.ea_ModellingAndVerificationOfSensors:2018}.
Finally, \emph{plato-graphical} models \cite{birkedal} use three perspectives: \emph{context} for   a true environment; \emph{proxy} as  a representation of the context (the shadows on the wall); and \emph{agents} that interact with the real context (environment) through the proxy.

\subsection{Example: Multi-perspective modelling of a building}\label{sec:multi_example}

So far, our building model has been  concerned mainly with the physical location of people within the building and their interactions with physical devices;   we might consider this a \emph{physical} perspective.

The people themselves have been kept abstract, \ie we have a single \biginline{Person} entity (ignoring Adult/Child distinction), although in practice people will have different \emph{roles} in the building, and relationships with others.
We could add these directly into the existing model, perhaps by nesting specific roles, \eg \biginline{Manager}, \biginline{Guard}, under a \biginline{Person} entity, but this mixes physical and social information.
In the multi-perspective approach we can instead introduce a  \emph{social} perspective  to track roles and relationships.

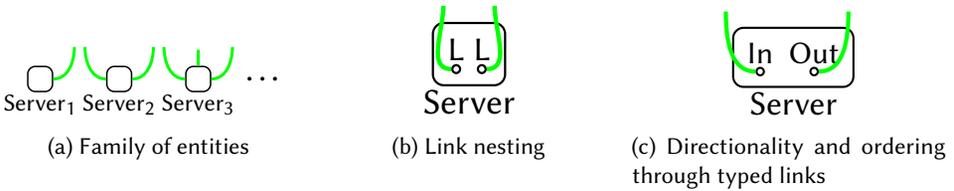
\begin{figure}
  \begin{subfigure}[b]{\linewidth}
    \centering
    \resizebox{0.7\linewidth}{!}{
       \begin{tikzpicture}[
social/.style = {draw},
physical/.style = {draw},
person/.style = {inner sep=0.2},
manages/.style = {draw},
manager/.style = {draw},
guard/.style = {draw},
employee/.style = {draw} ]
\node[person, label={[inner sep=0.5, name=n2l]north:{}}] (n2) {\Ladiesroom};
\node[server, right=0.3 of n2, label={[inner sep=0.5, name=n3l]north:{}}] (n3) {};
\node[person, right=0.4 of n3, label={[inner sep=0.5, name=n5l]north:{}}] (n5) {\Ladiesroom};
\node[room, right=0.5 of n5, label={[inner sep=0.5, name=n6l]north:{}}] (n6) {};
\node[guard, right=1.000000 of n6, label={[inner sep=0.5, name=n9l]north:{\sf\tiny Guard}}] (n9) {};
\node[manager, right=1.000000 of n9, label={[inner sep=0.5, name=n11l]north:{\sf\tiny Manager}}] (n11) {};
\node[manages, right=0.7 of n11, label={[inner sep=0.5, name=n12l]north:{\sf\tiny Manages}}] (n12) {};
\node[room, fit=(n3)(n3l)(n2)(n2l), label={[inner sep=0.5, name=n1l]north:{}}] (n1) {};
\node[room, fit=(n5)(n5l), label={[inner sep=0.5, name=n4l]north:{}}] (n4) {};
\node[employee, fit=(n9)(n9l), label={[inner sep=0.5, name=n8l]north:{\sf\tiny Employee}}] (n8) {};
\node[employee, fit=(n12)(n12l)(n11)(n11l), label={[inner sep=0.5, name=n10l]north:{\sf\tiny Employee}}] (n10) {};
\node[physical, fit=(n6)(n6l)(n4)(n4l)(n1)(n1l), label={[inner sep=0.5, name=n0l]north:{\sf\tiny Physical}}] (n0) {};
\node[social, fit=(n10)(n10l)(n8)(n8l), label={[inner sep=0.5, name=n7l]north:{\sf\tiny Social}}] (n7) {};
\node[big region, fit=(n0)(n0l)] (r0) {};
\node[big region, fit=(n7)(n7l)] (r1) {};
\node[yshift=40, xshift=50] (name_s1) {\tiny $s1$};
\node[right=1 of name_s1] (name_s2) {\tiny $s2$};
\draw[big edge] (n2) to[out=-90,in=-90] (n3);
\draw[big edgec] (n5) to[out=180,in=-90] ($(n5) + (-0.3,0.3)$);
\coordinate (h0) at ($(n2) + (1,1)$);
\draw[big edge, red] (n2) to[out=0,in=180] (h0);
\draw[big edge, red] (n8) to[out=180,in=0] (h0);
\draw[big edge, red] (n12) to[out=180,in=0] (h0);
\draw[big edge, red] (name_s1) to[out=-90,in=90] (h0);
\coordinate (h1) at ($(n5) + (1,1)$);
\draw[big edge] (n5) to[out=0,in=180] (h1);
\draw[big edge] (n10) to[out=180,in=0] (h1);
\draw[big edge] (name_s2) to[out=-90,in=90] (h1);
  \end{tikzpicture}
     }
    \caption{}
    \label{fig:multiperspective_state}
  \end{subfigure}

  \begin{subfigure}[b]{\linewidth}
    \centering
\begin{bigrapher}
# Perspectives
ctrl Physical = 0;
ctrl Social = 0;

# Physical
ctrl Room = 0;
atomic ctrl Server = 1;
atomic ctrl Person = 2;

# Social
ctrl Employee = 1;
atomic ctrl Manager = 0;
atomic ctrl Manages = 1;
atomic ctrl Guard = 0;

big multipersp =
    Physical.(/x Room.(Person{s1,x} | Server{x}) | /y Room.Person{s2,y} | Room.1)
 || Social.(Employee{s1}.Guard | Employee{s2}.(Manager | Manages{s1}));
\end{bigrapher}
    \caption{}
    \label{fig:multiperspective_big}

  \end{subfigure}
  \caption{Building model with two perspectives: Physical and Social. (a) Bigraph model, coloured links distinguish hyperedges (b) BigraphER snippet.}
  \label{fig:multiperspective_example}
\end{figure}

\Cref{fig:multiperspective_example}
 shows a simple multi-perspective model of a building extended with a social perspective.
To enable cross-perspective linking, we have added an additional link to the \biginline{Person} entity  (alternatively we could      introduce the link to a nested entity:    
see \cref{sec:fixed_arity}).
To make it clear we colour these links differently per person.
For each \biginline{Person} in the physical perspective, they have a social representation as an \biginline{Employee} (assuming no visitors are allowed in this building).
Not all entities will have a representation in other perspectives, \eg the \biginline{Servers}   exist physically but not socially.
Within the social perspective, employees have information about roles and possible perspective-local links, \eg to represent management organisation.

\tipbox{For multi-perspective modelling it is useful to add a top-level entity that allows a region to be referred to by name, \eg \biginline{Physical}, \biginline{Social}. }

The multi-perspective approach is extensible.
For example, and additional perspective could model an employee (virtual) staff record, \ie a \emph{Database} perspective, by linking to the \biginline{Employee} in the \emph{Social} perspective.
Not only does this mean the database can reflect social changes, but it can also reflect physical changes directly as all three representations share a hyperlink.

\section{More entities or more links? } \label{sec:practical}

While the place and link graph are   disjoint relations, the interactions between them can be used to model a wide  range of scenarios. The general advice is as follows:

\tipbox{If it is difficult to model with places, add more link structure; if it is difficult to model with links, add more place structure.}

In the next four  sections we give examples of  adding  place structure to increase expressiveness.

\subsection{Overcoming Fixed Arity}
\label{sec:fixed_arity}

\begin{figure}
  \centering
	\begin{subfigure}[t]{0.3\linewidth}
    \centering
    \resizebox{\linewidth}{!}{
      \begin{tikzpicture}
  \node[server, label={[inner sep=0.5]south:{\tiny $\mathsf{Server_1}$}}] (s1) {};
  \draw[big edge] (s1.east) to[out=0, in=-90] ($(s1.east) + (0.2,0.3)$);
  \node[server, right=0.5 of s1, label={[inner sep=0.5]south:{\tiny $\mathsf{Server_2}$}}] (s2) {};
  \draw[big edge] (s2.east) to[out=0, in=-90] ($(s2.east) + (0.2,0.3)$);
  \draw[big edge] (s2.west) to[out=180, in=-90] ($(s2.west) + (-0.2,0.3)$);
  \node[server, right=0.5 of s2, label={[inner sep=0.5]south:{\tiny $\mathsf{Server_3}$}}] (s3) {};
  \draw[big edge] (s3.east) to[out=0, in=-90] ($(s3.east) + (0.2,0.3)$);
  \draw[big edge] (s3.west) to[out=180, in=-90] ($(s3.west) + (-0.2,0.3)$);
  \draw[big edge] (s3.north) to[out=90, in=-90] ($(s3.north) + (0,0.15)$);
  \node[right=0.2 of s3] (dots) {\dots};
\end{tikzpicture}
     }
    \caption{Family of entities}
    \label{D}
  \end{subfigure}
	\begin{subfigure}[t]{0.3\linewidth}
    \centering
    \resizebox{0.4\linewidth}{!}{
      \begin{tikzpicture}
  \node[draw, circle, inner sep=0.5, label={[inner sep=0.5, name=l1l]north:{\tiny\sf L}}](l1) {};
  \node[draw, circle, inner sep=0.5, label={[inner sep=0.5, name=l2l]north:{\tiny\sf L}}, right=0.1 of l1] (l2) {};
  \node[server, fit=(l1)(l1l)(l2)(l2l), inner sep=2, label={[inner sep=0.5]south:{\tiny $\mathsf{Server}$}}] (s1) {};

  \draw[big edge] (l2.east) to[out=0, in=-90] ($(l2.east) + (0.05,0.4)$);
  \draw[big edge] (l1.west) to[out=180, in=-90] ($(l1.west) + (-0.05,0.4)$);
\end{tikzpicture}
     }
    \caption{Link nesting}
    \label{fig:rr_fixed_arity_nesting}
  \end{subfigure}
	\begin{subfigure}[t]{0.3\linewidth}
    \centering
    \resizebox{0.5\linewidth}{!}{
      \begin{tikzpicture}
  \node[draw, circle, inner sep=0.5, label={[inner sep=0.5, name=l1l]north:{\tiny\sf In}}](l1) {};
  \node[draw, circle, inner sep=0.5, label={[inner sep=0.5, name=l2l]north:{\tiny\sf Out}}, right=0.3 of l1] (l2) {};
  \node[server, fit=(l1)(l1l)(l2)(l2l), inner sep=2, label={[inner sep=0.5]south:{\tiny $\mathsf{Server}$}}] (s1) {};

  \draw[big edge] (l2.east) to[out=0, in=-90] ($(l2.east) + (0.2,0.4)$);
  \draw[big edge] (l1.west) to[out=180, in=-90] ($(l1.west) + (-0.2,0.4)$);
\end{tikzpicture}
     }
    \caption{Directionality and ordering through typed links}
    \label{fig:rr_fixed_arity_order}
  \end{subfigure}
  
  \begin{subfigure}[b]{\linewidth}
    \centering
\begin{bigrapher}
ctrl Server = 0;
atomic ctrl L = 1; # Links

big tree_network = Server.(L{x} | L{y}) || Server.L{x} || Server.L{y};
\end{bigrapher}
    \caption{}
    \label{fig:fixed_arity_big}

  \end{subfigure}
  \caption{Overcoming fixed arity and port ordering. a)   The family of \biginline{Server} entities   with 1, 2, 3 etc. ports   we want to represent. b)   The new definition of \biginline{Server} where instead of \eg 2 links, we have 2 nested \biginline{L}  entities.
  c) New entities  \biginline{In} and \biginline{Out}.}
  \label{fig:fixed_arity}
\end{figure}

Recall that  entities  have  fixed arity, \ie  a fixed  number of ports.   
   A  common   modelling  paradigm  is some objects  have a varying number of ports, \eg a server that may have $n$ (point to point) connections.

Intuitively, we would like   to model this with  a \emph{family} of \biginline{Server} entities, \ie $\mathsf{Server_{n}}$ for $0 \le n \le \ {max\_connections}$, where each $\mathsf{Server_n}$ allows  $n$  links. In effect, these are   sub-types, which are not   supported in  bigraphs. So, each of these entities would have to be specified separately,    requiring an explosion in the number of rewrite rules required. 

A practical way to deal with this  is to   introduce a new entity of arity 1 that represents a link endpoint. Now, whenever we need to add a new connection, we   simply nest a new \biginline{L} entity and use that link to form the connection. This reduces the total number of entities required  to 2 instead of a family of $n$, and there is now no problem to match directly on a \biginline{Server} entity regardless of the connections (by hiding them in a site). This approach is shown in \cref{D}  and
 \cref{fig:rr_fixed_arity_nesting}
        and is an example of    using the \emph{place} graph to fix what is essentially an  expressiveness of  \emph{link} problem. \Cref{fig:fixed_arity_big} shows BigraphER code using this technique to model a simple tree network topology.

\subsection{Directed Links }
\label{sec:dir_links}

It is possible to use this   approach to encode directed links.
For example, instead of introducing a single \biginline{L} entity, we introduce a pair of entities  \biginline{Out} and \biginline{In} (illustrated in  \cref{fig:rr_fixed_arity_order}) entities representing the source and target of a link resp.

This approach should not be confused with directed bigraphs~\cite{grohmann.ea_DirectedBigraphs:2007}, an extension to the bigraph theory where the link graph itself is directed. While they both allow direction to be specified, directed bigraphs allows \emph{names} to also have direction\footnote{Directed bigraphs are designed to allow bisimulation congruences to be derived in the face of name aliasing rather than  to solve specifically the directed link problem in models.}.

\subsection{Typed Ports}
\label{sec:typed_ports}

Recall also that entities have  \emph{unordered} ports, meaning that an entity such as \biginline{A\{input, out\}}
does not guarantee that the first port connects to input---because there is no such thing as a \emph{first} port!

To gain control over the ports, we can use the same trick as above: introduce an entity for each type of port we need, \eg \biginline{In\{i\}}, \biginline{Out\{o\}}, thus making it unambiguous which  link we want to use.
The ports themselves remain the same, instead we add ordering information through more structure.
This is shown in \cref{fig:rr_fixed_arity_order}.

\subsection{Ordered children}
\label{sec:ordered_children}

Similarly, the children of a entity are \emph{unordered}, and merge product is commutative.
If we want to   order  children we can again add extra entities.   For example, \textsf{L} and \textsf{R} might represent the left and right argument of a subtraction function, \eg \biginline{Sub.(L.Int(5) | R.Int(2))} meaning $5-2$.

To avoid introducing an entire family of entities when working with long lists of ordered arguments, \eg \biginline{Arg_1}, \biginline{Arg_2}, \dots, we can encode a linked-list structure, the simplest being a single \biginline{Cons} entity and 1 to represent the empty cell.
For example \biginline{b | Cons.(b' | Cons.(b'' | 1))} is 
a list of length 3 (where \biginline{b},\biginline{b'},\dots are arbitrary bigraphs). 
Alternatively a list could be encoded through links, \eg with links representing \emph{pointers} to the next cell.
In both cases it is possible to use rewrite rules  to implement the usual functional abstractions for lists: map, fold \etc.

\section{Applying a rule a fixed number of times and taking turns} \label{sec:tagging}

\begin{figure}
  \centering
	\begin{subfigure}[b]{\linewidth}
		\centering
		
    \resizebox{0.6\linewidth}{!}{
      \begin{tikzpicture}[
  vault/.append style = {draw},
  loginT/.append style = {draw},
  login/.append style = {draw},
  closed/.append style = {draw} ]
  \begin{scope}[local bounding box=lhs, shift={(0,0)}]
    \node[closed,  label={[inner sep=0.5, name=n1l]north:{\sf\tiny Closed}}] (n1) {};
    \node[vault, fit=(n1)(n1l), label={[inner sep=0.5, name=n0l]north:{\sf\tiny Vault}}] (n0) {};
    \node[big region, fit=(n0)(n0l)] (r0) {};
  \end{scope}
  \begin{scope}[local bounding box=rhs, shift={(2,0)}]
    \node[closed,  label={[inner sep=0.5, name=n1l]north:{\sf\tiny Closed}}] (n1) {};
    \node[login, right=0.6 of n1, label={[inner sep=0.5, name=n2l]north:{\sf\tiny Login}}] (n2) {};
    \node[loginT, right=0.6 of n2, label={[inner sep=0.5, name=n3l]north:{\sf\tiny LoginT}}] (n3) {};
    \node[loginT, right=0.6 of n3, label={[inner sep=0.5, name=n4l]north:{\sf\tiny LoginT}}] (n4) {};
    \node[vault, fit=(n4)(n4l)(n3)(n3l)(n2)(n2l)(n1)(n1l), label={[inner sep=0.5, name=n0l]north:{\sf\tiny Vault}}] (n0) {};
    \node[big region, fit=(n0)(n0l)] (r0) {};
  \end{scope}

  \node[] at ($($(lhs.east)!0.5!(rhs.west)$) + (0,0)$) (rule) {$\rrul$};
\end{tikzpicture}
     }
    
    \resizebox{0.5\linewidth}{!}{
      \begin{tikzpicture}[
  vault/.append style = {draw},
  person/.append style = {draw},
  loginT/.append style = {draw} ]
  \begin{scope}[local bounding box=lhs, shift={(0,0)}]
    \node[person,  label={[inner sep=0.5, name=n0l]north:{\sf\tiny Person}}] (n0) {};
    \node[loginT, right=0.6 of n0, label={[inner sep=0.5, name=n2l]north:{\sf\tiny LoginT}}] (n2) {};
    \node[big site, right=0.4 of n2,] (s0){};
    \node[vault, fit=(n2)(n2l)(s0), label={[inner sep=0.5, name=n1l]north:{\sf\tiny Vault}}] (n1) {};
    \node[big region, fit=(n1)(n1l)(n0)(n0l)] (r0) {};

  \end{scope}
  \begin{scope}[local bounding box=rhs, shift={(3.5,0)}]
    \node[loginT,  label={[inner sep=0.5, name=n1l]north:{\sf\tiny LoginT}}] (n1) {};
    \node[big site, right=0.6 of n1,] (s0){};
    \node[vault, fit=(s0), label={[inner sep=0.5, name=n2l]north:{\sf\tiny Vault}}] (n2) {};
    \node[person, fit=(n1)(n1l), label={[inner sep=0.5, name=n0l]north:{\sf\tiny Person}}] (n0) {};
    \node[big region, fit=(n2)(n2l)(n0)(n0l)] (r0) {};

  \end{scope}

  \node[] at ($($(lhs.east)!0.5!(rhs.west)$) + (0,0)$) (rule) {$\rrul$};
\end{tikzpicture}
     }
    
    \resizebox{0.4\linewidth}{!}{
      \begin{tikzpicture}[
  vault/.append style = {draw},
  login/.append style = {draw},
  open/.append style = {draw},
  closed/.append style = {draw} ]
  \begin{scope}[local bounding box=lhs, shift={(0,0)}]
    \node[closed,  label={[inner sep=0.5, name=n1l]north:{\sf\tiny Closed}}] (n1) {};
    \node[login, right=0.6 of n1, label={[inner sep=0.5, name=n2l]north:{\sf\tiny Login}}] (n2) {};
    \node[vault, fit=(n2)(n2l)(n1)(n1l), label={[inner sep=0.5, name=n0l]north:{\sf\tiny Vault}}] (n0) {};
    \node[big region, fit=(n0)(n0l)] (r0) {};
  \end{scope}
  \begin{scope}[local bounding box=rhs, shift={(3,0)}]
    \node[open,  label={[inner sep=0.5, name=n1l]north:{\sf\tiny Open}}] (n1) {};
    \node[vault, fit=(n1)(n1l), label={[inner sep=0.5, name=n0l]north:{\sf\tiny Vault}}] (n0) {};
    \node[big region, fit=(n0)(n0l)] (r0) {};
  \end{scope}
  \node[] at ($($(lhs.east)!0.5!(rhs.west)$) + (0,0)$) (rule) {$\rrul$};
\end{tikzpicture}
     }
    
    \resizebox{0.4\linewidth}{!}{
      \begin{tikzpicture}[
  vault/.append style = {draw},
  login/.append style = {draw},
  closed/.append style = {draw} ]
  \begin{scope}[local bounding box=lhs, shift={(0,0)}]
    \node[closed,  label={[inner sep=0.5, name=n1l]north:{\sf\tiny Closed}}] (n1) {};
    \node[login, right=0.6 of n1, label={[inner sep=0.5, name=n2l]north:{\sf\tiny Login}}] (n2) {};
    \node[big site, right=0.4 of n2,] (s0){};
    \node[vault, fit=(n2)(n2l)(n1)(n1l)(s0), label={[inner sep=0.5, name=n0l]north:{\sf\tiny Vault}}] (n0) {};
    \node[big region, fit=(n0)(n0l)] (r0) {};

  \end{scope}
  \begin{scope}[local bounding box=rhs, shift={(3.5,0)}]
    \node[closed,  label={[inner sep=0.5, name=n1l]north:{\sf\tiny Closed}}] (n1) {};
    \node[vault, fit=(n1)(n1l), label={[inner sep=0.5, name=n0l]north:{\sf\tiny Vault}}] (n0) {};
    \node[big region, fit=(n0)(n0l)] (r0) {};

  \end{scope}

  \node[] at ($($(lhs.east)!0.5!(rhs.west)$) + (0,0)$) (rule) {$\rrul$};
\end{tikzpicture}
     }
    
    \resizebox{0.4\linewidth}{!}{
      \begin{tikzpicture}[
  vault/.append style = {draw},
  person/.append style = {draw},
  loginT/.append style = {draw} ]
  \begin{scope}[local bounding box=lhs, shift={(0,0)}]
    \node[big site, ] (s0){};
    \node[loginT, right=0.6 of s0, label={[inner sep=0.5, name=n2l]north:{\sf\tiny LoginT}}] (n2) {};
    \node[vault, fit=(s0), label={[inner sep=0.5, name=n0l]north:{\sf\tiny Vault}}] (n0) {};
    \node[person, fit=(n2)(n2l), label={[inner sep=0.5, name=n1l]north:{\sf\tiny Person}}] (n1) {};
    \node[big region, fit=(n1)(n1l)(n0)(n0l)] (r0) {};
  \end{scope}
  \begin{scope}[local bounding box=rhs, shift={(3,0)}]
    \node[big site, ] (s0){};
    \node[person, right=0.6 of s0, label={[inner sep=0.5, name=n1l]north:{\sf\tiny Person}}] (n1) {};
    \node[vault, fit=(s0), label={[inner sep=0.5, name=n0l]north:{\sf\tiny Vault}}] (n0) {};
    \node[big region, fit=(n1)(n1l)(n0)(n0l)] (r0) {};
  \end{scope}

  \node[] at ($($(lhs.east)!0.5!(rhs.west)$) + (0,0)$) (rule) {$\rrul$};
\end{tikzpicture}
     }
	\caption{\biginline{tryOpen}, \biginline{login}, {\biginline{open}, \biginline{failed}, and \biginline{clean} rewrite rules.}}
	\end{subfigure}

	\begin{subfigure}[b]{\linewidth}
		\centering
    \begin{bigrapher}
ctrl Vault = 0;
ctrl Person = 0;

atomic ctrl Login = 0;
atomic ctrl LoginT = 0;
atomic ctrl Closed = 0;
atomic ctrl Open = 0;

# start
react tryOpen =
  Vault.Closed --> Vault.(Closed | Login |  LoginT | LoginT);

# apply
react login =
  Person.1 | Vault.(LoginT | id) --> Person.LoginT | Vault.id;

# Base cases
react open =
  Vault.(Closed | Login) --> Vault.Open;

react failed =
  Vault.(Closed | Login | id) --> Vault.Closed @[];

# Cleanup
react clean =
  Vault.id | Person.LoginT --> Vault.id | Person.1 if !(Login) in param;

begin brs
  init ...;
  rules = [ {clean}, { tryOpen, login, open}, {failed} ];
end
    \end{bigrapher}
		\caption{}
	\end{subfigure}
  \caption{Tagging example: vault login process. Two people are required to open the vault. }
  \label{fig:tagging}
\end{figure}

Often we require to apply a   rule, or a sequence of rules,  exactly $n$ times, which is not expressible within  a  standard rewriting framework.  
 In bigraphs, we can encode   rule control  by  \emph{tagging}: using additional \emph{tag} entities  to mark where rules have previously been applied.
Note that tagging    requires  rule priorities (\cref{sec:priorities}) or conditional rules (\cref{sec:conditionals}).

 We illustrate with an example. 
 Consider the scenario     where access to the vault of a building requires  $n$  people to log into the vault access system at the same time. The model   is in  \cref{fig:tagging}. Essentially,  
the vault is initially  \biginline{Closed}, then $n$ people 
 \biginline{Login}, and then the vault is \biginline{Open}.
But,  {\em any} number of people may be {\em trying} to gain access and so    we have to restrict  successful logins to exactly    $n$. 
 In other words, we cannot allow the rule for \biginline{login} 
  to be applied  any
number of times. 
 For simplicity we assume $n=2$ in this example.

We introduce two new tags   for controlling the reaction sequence: \biginline{Login}   denotes a login sequence has started and \biginline{LoginT}   controls the number of people required to open the vault---we use it to   distinguish those that have already logged in from those that have not.
The rules are modified to reflect   four phases:

\begin{description}
  \item[\textbf{Start tagging}]    
  Add a tag to  denote the sequence has started and is in progress. In this case \rr{tryOpen} adds an entity \biginline{Login} to the \biginline{Vault} to note the start of a login sequence. As we want a fixed number of \rr{login}s, we additionally add $n$ \biginline{LoginT} tokens (in this case 2, but can be changed without affecting the rest of the login mechanism).

  \item[\textbf{Apply rule(s)}] Once the sequence has started,   apply the rule(s) $n$ times. 
In this case \rr{login} allows a \biginline{Person}, in the same room as the vault (note use of \biginline{|} and not  \biginline{||}) who has not logged in before, \ie does not nest a \biginline{LoginT} tag, to perform a login by accepting the \biginline{LoginT} token.

  \item[\textbf{Base case(s)}] The base case determines when the sequence of operations should stop, \eg when there are no matches left. In this case \rr{open} checks all \biginline{LoginT} tokens have been taken, allowing the vault door to open. If not enough people  log in, a second base case \rr{failed} stops the login sequence. Rule priorities are used to enforce \rr{failed} can only be applied once we have checked all possible applications of \rr{login}. 

  \item[\textbf{Cleanup}] The final step   removes redundant tags. Here, the conditional rule  \rr{clean} removes \biginline{LoginT} entities   when the \biginline{Login} sequence has ended (either successfully or with failure). 
\end{description}

For tools that do not support conditional bigraphs, we can add an extra tag, \eg \biginline{LoginDone}, to determine when the sequence has ended in order to control when cleanup can happen.

\tipbox{If you want to apply a rule sequence  a fixed number of times, add tags  to entities to indicate whether or not the sequence has been applied, and modify    the  rules so they  introduce and then remove tags.}

While powerful, tagging   increases the number of entities and can make it less clear how entities are supposed to be used. One solution is to use tagging   with instantaneous rules, see   \cref{sec:instantaneous},   this allows us to apply sequences transparently, as if in a single step. An alternative approach is to define a family of rules for each number of people we need to log in, \eg \rr{open\_vault\_1}, \rr{open\_vault\_2},~\dots, but as with parameterised rules, this can lead to an explosion of the total number of rules.

\subsection{Turn taking or phases of operation}
\label{sec:turntaking}

A 
similar, but distinct scenario is  modelling   \emph{phases} of operation.
For example, we might have a \emph{movement} phase, where all people in our building can (but don't have to) move between rooms.
Once \emph{everyone} has made a movement (or idled), we might then have a \emph{sensing} phase, where the security cameras try to detect intruders, before moving back to the movement phase.

This sort of turn-taking is well captured using multi-perspective modelling (\cref{sec:multiperspective}), where a \emph{Control} perspective can track the current state of the system, \eg \emph{Movement} vs \emph{Sensing}.
As perspectives are just additional parallel regions, it is   easy to match on them, \ie we just extend a rule $L \rrul R$ to $L \parallel C \rrul R \parallel C$ for some control information encoded by $C$.

Alternatively we can use a conditional rule (\cref{sec:conditionals}, $L \rrul R\;$ \biginline{if C in ctx} If \biginline{C} does not change during rule execution.
Keeping track of who still needs to move/sense can be done by nesting additional tokens, \eg \biginline{Person.Move}, \biginline{Person.Sense}.

A snippet of a movement and sensing example is   in \cref{fig:turntaking}.
We show both direct matches and conditionals; in general it is good practice to use a single style only.
 
Key to turn taking is phase shift functions that move between movement and sensing, \eg they model a state machine.
Here we have used context conditions (\cref{sec:conditionals}) to check   all \biginline{Person} entities have acted before swapping phase.
If conditionals are not supported, then tagging (\cref{sec:tagging}) could be used.

\tipbox{There may be several possible ways to model a system feature, \eg tagging, conditional rewriting, parameterisation, \etc  it is good practice to employ one consistent approach throughout the model.}

\begin{figure}
    \begin{bigrapher}
ctrl Room = 0;

ctrl Person = 0;
atomic ctrl Move = 0;
atomic ctrl Sense = 0;

atomic ctrl Camera = 0;
atomic ctrl Alarm = 0;

# Phases
ctrl Control = 0;
atomic ctrl Movement = 0;
atomic ctrl Sensing = 0;

# Matching style
react move =
  Room.(id | Person.Move) || Room.id || Control.Movement
  -->
  Room.id || Room.(id | Person.Sense) || Control.Movement;

# Conditional Style
react sense =
  Room.(id | Camera | Person.Sense)
  -->
  Room.(id | Camera | Person.Move | Alarm)
  if Control.Sensing in ctx;

react no_sense =
  Room.(id | Person.Sense)
  -->
  Room.(id | Person.Move)
  if Control.Sensing in ctx, !Camera in param;

# Phase shifts
react move_sense = Control.Movement --> Control.Sensing if !Person.Move in ctx;

react sense_move = Control.Sensing --> Control.Movement if !Person.Sense in ctx;
    \end{bigrapher}
  \caption{Turn Taking: BigraphER snippet for movement and sensing phases.}
  \label{fig:turntaking}
\end{figure}

\section{Further extensions: probabilistic,   stochastic,   and non-deterministic rewriting}
\label{sec:BRS2}
In standard BRSs, reaction rules are applied non-deterministically: {\em any} rule (within the current priority class) that has a match in the current state can be applied. Selection is    random. Extensions to BRSs allow reaction rules to be annotated, allowing them to be applied probabilistically~\cite{archibald_ProbabilisticBigraphs:2021}, stochastically (exiting a state at some rate)~\cite{DBLP:journals/entcs/KrivineMT08}, or through explicit non-deterministic action choice, in the style of an Markov Decision Process~\cite{archibald_ProbabilisticBigraphs:2021}.

For each extension, we annotate a rule $r : L \rrul R$ with additional information, \eg $r_{p} : L \rrulp{2} R$ is a probabilistic rule with \emph{weight} 2; $r_{s} : L \rrulp{0.2} R$ is a stochastic rule with exit rate $0.2$; and $r_{a} : L \rrula{act}{4}$ is a rule with weight 4 if action $act$ is chosen.
Weights give the relative application chance for a rule.
That is for rules $r_{p} : L \rrulp{2} R$ and $r_{q} :  L \rrulp{4} R$, if both have a match in the current state, then $r_{q}$ should be twice ($\frac{4}{2}$) as likely to be applied.
Weightings are scaled relative to the number of matches possible, for example if $r_{p}$ has two valid matches, while $r_{q}$ only has one, then they are applied with the same probability.

An example probabilistic BRS is in \cref{fig:prob_example}.
A probabilistic BRS is requested using \biginline{begin pbrs} instead of \biginline{begin brs} in the BRS definition block.
In this example, we want to detect possible security threats using the security camera(s) in a room.
In practice sensors, \eg cameras, are not 100\% accurate, and we want to model this uncertainty.
As is common in probabilistic modelling, we use two rules representing the   cases \emph{detect}
and the converse   \emph{avoid\_detect}.
The weights give the relative probability of application -- in this case detection is 4 times as likely as avoidance. When one \biginline{Intruder} is in the room \rr{detect} applies (after normalisation) with probability $0.8$. 
In practice, when there are many matches and priority ordering on rules \etc it becomes much harder to predict the rule probability by hand.

We give two short examples of featuring stochastic rules and non-deterministic {\em action} choice in \cref{fig:stochastic_example,fig:mdp_example}.

Stochastic rules are like probabilistic rules but with \emph{rates} (positive real numbers) that
  determine how often a particular rule should be applied.   Rules with higher rates occur more often, \eg in \cref{fig:stochastic_example} people exit the building more often than they enter.

Non-deterministic {\em action} choice is offered in 
action bigraphs, which, like probabilistic bigraphs, specify a \emph{weight} between the left and right of a rule, however in this case the set of rules that can be applied is affected by the action they are in.
Actions are specified in the \biginline{begin abrs} block using a set syntax, \eg \biginline{actions = [ move = \{move\_stay, move\_room\} ]} meaning that when move is chosen only reactions \biginline{move\_stay} and \biginline{move\_room} can be applied.
These rules are then applied in the same style as probabilistic bigraphs.

\tipbox{Probabilistic, stochastic, and action bigraphs control \emph{how} rules are applied, \eg how often, but do not affect the rules themselves that keep the same rewriting mechanism as before.}

\begin{figure}
  \centering
	\begin{subfigure}[b]{\linewidth}
		\centering
    \resizebox{0.6\linewidth}{!}{
      \begin{tikzpicture}[
room/.append style = {draw},
intruder/.append style = {draw},
camera/.append style = {draw},
alarm/.append style = {draw} ]
  \begin{scope}[local bounding box=lhs, shift={(0,0)}]
    \node[intruder, label={[inner sep=0.5, name=n1l]north:{\sf\tiny Intruder}}] (n1) {};
    \node[camera, right=0.6 of n1, label={[inner sep=0.5, name=n2l]north:{\sf\tiny Camera}}] (n2) {};
    \node[big site, right=0.4 of n2] (s1) {};
    \node[room, fit=(n2)(n2l)(n1)(n1l)(s1), label={[inner sep=0.5, name=n0l]north:{}}] (n0) {};
    \node[big region, fit=(n0)(n0l)] (r0) {};
  \end{scope}
  \begin{scope}[local bounding box=rhs, shift={(3.5,0)}]
    \node[intruder, label={[inner sep=0.5, name=n1l]north:{\sf\tiny Intruder}}] (n1) {};
    \node[camera, right=0.6 of n1, label={[inner sep=0.5, name=n2l]north:{\sf\tiny Camera}}] (n2) {};
    \node[alarm, right=0.6 of n2, label={[inner sep=0.5, name=n3l]north:{\sf\tiny Alarm}}] (n3) {};
    \node[big site, right=0.4 of n3] (s1) {};
    \node[room, fit=(n3)(n3l)(n2)(n2l)(n1)(n1l)(s1), label={[inner sep=0.5, name=n0l]north:{\sf\tiny Room}}] (n0) {};
    \node[big region, fit=(n0)(n0l)] (r0) {};
  \end{scope}

 \node[] at ($($(lhs.east)!0.5!(rhs.west)$) + (0,0)$) (rule) {$\rrulp{4}$};
\end{tikzpicture}
     }
    \caption{}
    \label{fig:rr_detect}
        
  \end{subfigure}
	\begin{subfigure}[b]{\linewidth}
		\centering
    \resizebox{0.6\linewidth}{!}{
      \begin{tikzpicture}[
room/.append style = {draw},
intruder/.append style = {draw},
camera/.append style = {draw}]
  \begin{scope}[local bounding box=lhs, shift={(0,0)}]
    \node[intruder, label={[inner sep=0.5, name=n1l]north:{\sf\tiny Intruder}}] (n1) {};
    \node[camera, right=0.6 of n1, label={[inner sep=0.5, name=n2l]north:{\sf\tiny Camera}}] (n2) {};
    \node[big site, right=0.4 of n2] (s1) {};
    \node[room, fit=(n2)(n2l)(n1)(n1l)(s1), label={[inner sep=0.5, name=n0l]north:{\sf\tiny Room}}] (n0) {};
    \node[big region, fit=(n0)(n0l)] (r0) {};
  \end{scope}
  \begin{scope}[local bounding box=rhs, shift={(3.5,0)}]
    \node[intruder, label={[inner sep=0.5, name=n1l]north:{\sf\tiny Intruder}}] (n1) {};
    \node[camera, right=0.6 of n1, label={[inner sep=0.5, name=n2l]north:{\sf\tiny Camera}}] (n2) {};
    \node[big site, right=0.4 of n2] (s1) {};
    \node[room, fit=(n2)(n2l)(n1)(n1l)(s1), label={[inner sep=0.5, name=n0l]north:{\sf\tiny Room}}] (n0) {};
    \node[big region, fit=(n0)(n0l)] (r0) {};
  \end{scope}

  \node[] at ($($(lhs.east)!0.5!(rhs.west)$) + (0,0)$) (rule) {$\rrulp{1}$};
\end{tikzpicture}
     }
    \caption{}
    \label{fig:rr_avoid_detect}
  \end{subfigure}

	\begin{subfigure}[b]{\linewidth}
		\centering
    \begin{bigrapher}
atomic ctrl Intruder = 0;
atomic ctrl Camera = 0;
atomic ctrl Alarm = 0;
ctrl Room = 0;

react detect =
  Room.(Intruder | Camera | id)
  -[4]->
  Room.(Intruder | Camera | Alarm | id);

react avoid_detect =
  Room.(Intruder | Camera | id)
  -[1]->
  Room.(Intruder | Camera | id);
  
begin pbrs
  init ...;
  rules = [ {detect, avoid_detect} ];
end
    \end{bigrapher}
		\caption{}
    \label{rr_prob_bigrapher}
	\end{subfigure}
  \caption{Probabilistic reaction rules with weights. (a)   Rule  \rr{detect} has weight 4. (b) Rule  \rr{avoid\_detect} has weight 1. (c)   BigraphER snippet.}
  \label{fig:prob_example}
\end{figure}

\begin{figure}
    \centering
		\centering
    \begin{bigrapher}
atomic ctrl Intruder = 0;
atomic ctrl Person = 0;
ctrl Room = 0;
ctrl Entrance = 0;

big s0 = Room.Entrance.1;

react enter = Room.Entrance.id -[0.2]-> Room.Entrance.(id | Person);

react exit =  Room.Entrance.(id | Person) -[0.3]-> Room.Entrance.id;

react enter_intruder = Room.Entrance.id -[0.01]-> Room.Entrance.(id | Intruder);

begin sbrs
  init s0;
  rules = [ {enter, exit, enter_intruder}];
end
    \end{bigrapher}
    \caption{Stochastic model of entrance hall: People exit the room more often than entering (rates of 0.3 vs 0.2). Intruders can enter but at a much lower rate (0.01).}
    \label{fig:stochastic_example}
\end{figure}

\begin{figure}
    \centering
		\centering
    \begin{bigrapher}
atomic ctrl Guard = 0;
atomic ctrl Intruder = 0;
ctrl Room = 0;
atomic ctrl Door = 1;
atomic ctrl Alarm = 0;

react move_stay = 
  Room.(id | Guard) -[5]-> Room.(id | Guard);
  
react move_room = 
  Room.(id | Door{x} | Guard) || Room.id 
  -[1]-> 
  Room.(id | Door{x}) || Room.(id | Guard);

react check_room = 
  Room.(id | Guard | Intruder) -[1]-> Room.(id | Alarm | Guard | Intruder); 
  
react check_room_safe = 
  Room.(id | Guard) -[1]-> Room.(id | Guard) if !Intruder in param; 
  
big s0 = /x (Room.(Door{x} | Guard) || Room.(Door{x} | Intruder));

begin abrs
  init s0;
  rules = [ {move_stay, move_room, check_room, check_room_safe}];
  actions = [ move = {move_stay, move_room}, check = {check_room, check_room_safe} ];
end
    \end{bigrapher}
    \caption{Action based (non-deterministic) model of guarded rooms. The system can decide to let guards move between rooms \emph{or} check the room they are in. Moving between rooms is less likely than  staying put (weight 5 vs weight 1). Only one of the two check rules is applicable when the action check is taken.}
    \label{fig:mdp_example}
\end{figure}

\section{Model  Analysis}
\label{sec:analysis}

While defining a BRS can be useful in its own right, \eg to document design decisions, we can perform analysis through  \emph{simulation} and
\emph{model checking}.  BigraphER   exports a transition system (a table of transitions; possibly labelled with probabilities/rates/actions), and an initial state, that can be   analysed in  tools such as    PRISM~\cite{Prism} or
STORM~\cite{Storm}.
The   initial  state is required to be   \emph{ground}---a bigraph that does not contain  sites or inner names. We specify the initial bigraph using \biginline{init b} (for some bigraph $b$), \eg line 26 of \cref{fig:model_checking_example}.
Before giving details of simulation and model checking, we introduce the concept of bigraph predicates, which we have found useful for analysis.

\subsection{Bigraph Patterns}
\label{sec:patterns}

Often   it is useful to label a subset of   states that have a (domain specific) feature of interest, \ie they they satisfy a {\em predicate}.   
 We specify the predicate    with a \emph{bigraph pattern}~\cite{benford.ea_Savannah:2016},  which abstracts the states that may match the left-hand-side    of a rewrite rule (the {\em pattern}). Bigraph patterns are simply (named) bigraphs that define static (state) properties.
 An example is in \cref{fig:model_checking_example}; 
 the patterns are indicated  by the reserved word \biginline{preds}
 (for {\em predicates}),   as shown on line 28.

A state that matches  pattern $p$   is then labelled with $p$,  and it may be labelled with more than one pattern.   
Pattern matching uses the same semantics as reaction rule matching, which 
  means we have  name equivalence, \ie \biginline{A\{x\}} and \biginline{A\{y\}} are the same pattern. 

  Bigraph  patterns may occur in   path formulae in temporal logics, an example using the patterns from \cref{fig:model_checking_example} is in 
  \cref{sec:model_checking}.

\subsection{Simulation}
\label{sec:sim}
 
In a simulated run  at each step we apply a single reaction rule (respecting priority and instantaneous rules), resulting in a single trace (of potentially many possible traces) through the system. Simulation is  particularly useful for  models with large state spaces.
BigraphER supports simulation for all extensions \eg probabilistic and stochastic rewriting.

To run a simulation of   \texttt{}{model.big} in BigraphER    use the following command line:
\begin{verbatim}
bigrapher sim -S <maxstates> -l <predicates.csl> <model.big>
\end{verbatim}

  where \texttt{predicates.csl} 
is the name of the (exported) file that     maps   bigraph patterns  to states.

\subsection{Model Checking}
\label{sec:model_checking}

In model checking we apply \emph{all} possible rules, in a breadth first manner, for a given state to explore  all possible traces.  Each new state is checked for equality with a previous state,   allowing loops in the underlying transition system.
BigraphER supports model checking for all extensions, \eg probabilistic, stochastic, and non-deterministic  rewriting by
exporting the transition system in     PRISM  format.\footnote{BigraphER natively supports PRISM format, but as the transition system is just a matrix it is likely to be supported by other tools with minor changes.}   
To generate the   transition system use the following command line:
{
\small
\begin{verbatim}
bigrapher full -M <maxstates> -l <predicates.csl> -p <transition.tra> <model.big>
\end{verbatim}
}
 
 where \texttt{predicates.csl}  and \texttt{transition.tra}  are the exported pattern mapping and transition system resp.    
Once the transition system is generated, we can   check  temporal logics properties expressed in logics such as     LTL, CTL, and PCTL~\cite{clarke1981design}.

\subsubsection{Example}

Servers   contain sensitive data and so server rooms should be secure. 
A simple   model   is in \cref{fig:model_checking_example}.
 \begin{figure}
\centering
\begin{bigrapher}
atomic ctrl Camera = 0;
atomic ctrl Server = 0;
atomic ctrl Intruder = 0;
ctrl Room = 0;
atomic ctrl Door = 1;
atomic ctrl Entrance = 0;

react move =
  Room.(Intruder | Door{x} | id) || Room.(id | Door{x})
  -->
  Room.(Door{x} | id) || Room.(Intruder | id | Door{x});

# Patterns/Predicates
big seen = Room.(Intruder | Camera | id);
big entrance = Room.(Entrance | Intruder | id);
big serverRoom = Room.(Server | Intruder | id);

big building = /d1/d2 (
   Room.(Entrance | Intruder | Door{d1})
|| Room.(Camera | Door{d1} | Door{d2})
|| Room.(Door{d1} | Door{d2})
|| Room.(Door{d2} | Server));

  
begin brs
   init building;
   rules = [{move}];
   preds = {seen, entrance, serverRoom};
end

\end{bigrapher}
  \caption{Secure building model: Can we reach a server without passing a camera?}
  \label{fig:model_checking_example}
\end{figure}

 The model has  explicit  \biginline{Door} entities that determine valid pairs of rooms.  
Assume that an  intruder is   captured by a camera---so long as they pass through a room containing  at least one camera. We   want to check whether there {\em is} an insecurity   via    a property such as: ``is  there a path to the server room that  passes only through  rooms that do not contain  a camera?''
We define three bigraph patterns: \emph{seen} that matches a room  containing at least one \biginline{Camera} and an \biginline{Intruder}; \emph{entrance} that  matches the entrance hall containing an \biginline{Intruder}; and  \biginline{serverRoom} that matches a room containing at least a \biginline{Server} and an \biginline{Intruder}.

\begin{figure}
  \centering
  \includegraphics[width=0.4\linewidth]{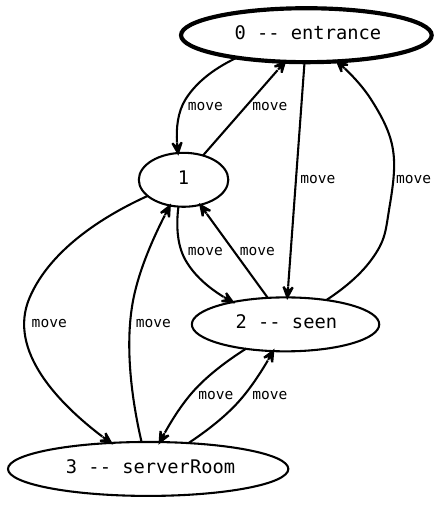}
  \caption{Transition system generated by the model of \cref{fig:model_checking_example}. Bold state is the starting state, numbers are state numbers, and labels are predicates.}
  \label{fig:secure_trans}
\end{figure}

The transition system generated by BigraphER is in \cref{fig:secure_trans}, and shows all possible movements between states.
As we do not model capture of the intruder they are always allowed to move even if they have been seen.
This means there are infinite traces possible, \eg $0 \to 1 \to 0 \to 1 \to 0 \to \dots$.

Although it is fairly obvious from the transition system that there is a path ($0 \to 1 \to 3$) to the server room without being seen, for larger models visually inspecting the transition system is error prone and likely too complex to  draw.
Instead, we can express the properties logically, in a form suitable for model checking software.
The property can be expressed by the (PRISM formatted) LTL formulae, using bigraph patterns, as:
\begin{verbatim}
E ["entrance" & (!"seen" U "serverRoom")]
\end{verbatim}

That is: there \emph{exists} (E) a path where initially   \texttt{entrance} holds, and for this state\footnote{This formula would not work if there was a camera in the entrance hall as seen would immediately be true.}, and all future states, \texttt{seen} does not hold \emph{until}  \texttt{serverRoom} becomes true.
The property holds as expected because  of the non-secured room (\ie a room without a camera)  between the entrance and the server room.

\subsection{Common Errors}

We end with a discussion of common errors you might see when executing your models with BigraphER and how to fix them.

\begin{description}
  \item[Init bigraph is not ground] When using models we assume the initial state is fully formed, \ie does not contain sites\footnote{or inner names.}. Usually this means you have a non-atomic entity with a (implicit) site: \biginline{big s0 = A;} (which means \biginline{big s0 = A.id}). \textbf{Fix} Put an empty bigraph in place of the site, \ie \biginline{big s0 = A.1}.

  \item[Invalid Reaction: Inner interfaces<1, \{\}> and <2, \{\}> do not match] This error means that the number of sites on the left hand side (1 in this case) does not match the number of sites on the right hand side (2 in this case), \ie we haven't specified what should go into the second site on the right. \textbf{Fix} When the number of sites are unequal you \emph{must} specify an instantiation map (\cref{sec:instantiation}) to determine the content of the sites after rewriting.

  \item[Invalid Reaction: Outer interfaces <1, \{x\}> and <1, \{\}> do not match] Outer interfaces (number of regions and outer names) are not allowed to change during a rewrite. In the example above we have lost the name $x$, \eg \biginline{A\{x\}.1 --> 1}. \textbf{Fix} if an open name is no longer required you should make it idle on the right of a rule, \eg \biginline{A\{x\}.1 --> \{x\} | 1} is the correct rule. A similar error occurs if you drop a parallel region, requiring additional \biginline{|| 1} components to fix.

  \item[Invalid Reaction: Instantiation map is not valid] This error means the instantiation map is badly formed. \textbf{Fix} ensure the map has an entry for each site on the right hand side and that entries point to valid sites on the left, \eg a map \biginline{@ [5]} when the left hand side only has one site is invalid.
\end{description}

\subsection{Further examples}
\label{sec:further_examples}
 We give an overview of five real-world systems we have modelled. For each, we  indicate if (and how) we employed multi-perspective modelling, sharing,  diagrammatic notation, rules, and analysis. With one exception, the shapes in the diagrammatic form are geometric (circle, rectangle, etc.). The exception is the mixed-reality game {\em Savannah} (\cref{Sav}) where  shapes that represent entities in the game,  such lions, impalas, and the human players,  are hand-drawn. 
 
The Bigraph code is available in the associated paper,  or     on the BigraphER website\footnote{https://uog-bigraph.bitbucket.io/examples.html},  which contains over 25 examples.

\subsubsection{ 802.11 CSMA/CA RTS/CTS communications protocol [10]}

\noindent This well-known communications  protocol   applies  to any network topology, including potentially overlapping wireless signals.  The protocol employs four way RTS/CTS  (ready to send/clear to send) handshaking. The system modelled includes arbitrary topologies and the possibility of   the hidden node problem – when two transmitting stations cannot sense each other, thus causing a collision, which is resolved through exponential backoff.   

\noindent
 {\em Multi-perspective modelling:} none.

\noindent 
 {\em  Sharing:}  for wireless signal ranges. 

\noindent \noindent
 {\em Diagrammatic notation:}  shapes for types, coloured shading  for different types or stage of data packets    (e.g. queued, sent,  CTS, RTS). 

\noindent
 {\em  Rules:}   priorities to implement instantaneous rules;  stochastic rules for rates of  transmission and retransmission. 

\noindent
  {\em Analysis: } model checking CSL (Continuous Stochastic Logic) properties  that express  quantitative, dynamic properties such as {\em what is  the probability of successful transmission of a packet} and  {\em what is the likelihood of a collision or being in an error state}.    

\subsubsection{ Mixed-reality multi-player game ~\cite {benford.ea_Savannah:2016}}\label{Sav} 

This ubiquitous system, called {\em Savannah}  (also mentioned in Section \ref{multi}), models the dynamic behaviour of players in a mixed-reality game in which the human players are instrumented    and their  real-world  physical location  affects  their capabilities  as players in the  game. The game involves wildlife and different types of terrain (the Savannah), which is mapped on to the physical playing space (\eg a field or football pitch).  There is different wildlife in each terrain.  The players hunt wildlife by forming and disbanding  teams; how and when they do that  depends on their proxemics: the personal (physical and social) space of each player.

\noindent
 {\em Multi-perspective modelling:} four perspectives --     Human, Physical, Computational, and Technology.

\noindent
 {\em  Sharing:}  for overlapping auras (personal space). 

\noindent
 {\em Diagrammatic notation:}  shapes for types,  coloured shading for  animals and players at different stages in the game  (e.g. idle lion, lion initiating an attack, lions and players in a group).  

\noindent
 {\em  Rules:}  the main model employs standard reaction rules, additionally,   there is  an  investigation of the use of  weights,  inferred from user trials,  to represent   GPS drift.

\noindent
  {\em Analysis: } simulation for replay of scenarios uncovered in user trials - scenarios in which some players experienced cognitive dissonance; manual inspection of rewrite rules for matches with  bigraph patterns, which revealed missing    relationships between some  perspectives and a system design flaw. 

\subsubsection{ Network and network policy  management~\cite{DBLP:journals/scp/CalderKSS14}} 

The management of a network  with an evolving topology, due to  network events such as   machines leaving and joining the network, is modelled.  In addition, the system has     dynamic  access control policies that  enforce or  forbid  certain behaviours  and thus constrain network evolutions.  Policies can be invoked or  lifted, as the network evolves, so event streams include both network and policy events.

\noindent
 {\em Multi-perspective modelling:} none. 

\noindent
 {\em  Sharing:}  for overlapping wireless signals.

\noindent 
 {\em Diagrammatic notation:}  shapes for types. 

\noindent
 {\em  Rules:}  standard reaction rules that make use     tagging to  model invocation  and lifting of  policies.

\noindent
  {\em Analysis: }  runtime  monitoring of   bigraph patterns that   express static properties such as {\em the current configuration complies with the current policies}. 
\subsubsection{ Large-scale sensor network infrastructures~\cite{sevegnani.ea_ModellingAndVerificationOfSensors:2018}} 

A model of sensor network  infrastructures  that enable investigations of   how  requirements for a sensor network can be  met, individually and collectively, and can continue to be met, in the context of large-scale, evolving network and device configurations. 
The exemplar is an urban sensor network infrastructure with two applications:  environmental monitoring and structural and  (material) health of buildings and bridges.

\noindent
 {\em Multi-perspective modelling:} three perspectives -- Physical, Data, and Services (requirements for applications).

\noindent
 {\em  Sharing:}  for overlapping wireless signals.

\noindent \noindent
 {\em Diagrammatic notation:}  shapes for types, coloured shading to distinguish  whether a node is in use  or in failure,  and for  different types of sensor.

\noindent
 {\em  Rules:}  standard reaction rules;   stochastic rules   for failure rates and repair rates.

\noindent
  {\em Analysis: } simulations using actual data streams and events generated by the Cooja network emulator/simulator~\cite{cooja}; runtime monitoring of 
   bigraph patterns that express   static properties such as 
   {\em there are sufficient nodes available in every network partition} and model checking 
  LTL (Linear Time Logic)  and CSL 
  (Continuous Stochastic Logic) properties that express   dynamic properties such as  
   {\em pollution, temperature, and humidity data are delivered when pollution levels exceed a threshold} and {\em  the probability of a node to fail within a certain time interval, while serving an app, is below a threshold.}
  
\subsubsection{CAN programming language for BDI agents~\cite{BDIBigraphs}} 

The operational   semantics for CAN (Conceptual Agent Notation),  an agent system  programming language, is encoded in bigraphs.  The structural encoding is natural, with bigraph reaction rules corresponding directly to the (semantics) inference rules. A motivation  for the encoding is to be able to verify  agent requirements.  The exemplar is a case-study based on UAV (unmanned aerial vehicles).

\noindent
 {\em Multi-perspective modelling:}   perspectives for B,D, and I -- Beliefs, Desires, Intentions, and for Plans.

\noindent
 {\em  Sharing:}  none.

\noindent
 {\em Diagrammatic notation:}    shapes for types, coloured shading for plan success or failure.

\noindent
 {\em  Rules:}  parameterised entities  and rules,   conditional rules, and rule priorities.

\noindent 
  {\em Analysis: }  
 bigraph patterns to express static properties such as {\em there is goal corresponding a given task} and  model checking CTL properties dynamic properties such as {\em the goal corresponding to a given task is persistent} and 
{\em two sensing tasks can always be completed regardless of their interleaving}.

\subsection{When to use Bigraphs}
\label{sec:why_bigraphs}

We have shown bigraphs to be an expressive modelling approach applicable to a wide range of scenarios such as the examples in \cref{sec:further_examples}.
While they are universal,  they are not always the best tool.
In this section we comment on the types of problems we think are best suited to Bigraphs, but you should feel free to experiment.

Bigraphs are based around \emph{relationships}, both spatial and non-spatial, between entities and are well suited to model situations where relationships change over time. 
For example, the networking examples of \cref{sec:further_examples} often allow dynamic connectivity between nodes.
Bigraphs are often much less suited for models with large amounts of state, and these are often better described with tools like Event-B~\cite{abrial_EventBBook:2010} that describe how states evolve over time.

While we support parameterised entities in BigraphER (\cref{sec:parameterised}) to enable some basic support for primitive data-types, \eg integers, bigraphs are not well suited to data-intensive applications that require, for example, filtering of a system based on values.
They are much more suited to symbolic analysis data, \eg showing properties of commutativity of data-types.

Bigraphs are primarily a \emph{system} design tool that determines how a design may respond to particular inputs/environments.
Given our analysis is largely based on model checking, not proof, bigraphs are not currently a good tool for \emph{code verification} as found in, for example, Isabelle/HOL\cite{nipkow.ea_IsabelleHolBook:2002}, or Lean\cite{mendon_ca_de_moura.ea_LeanDescription:2015}.
Similar system design is popular with tools such as TLA$^{+}$~\cite{lamport_TLAPLusBook:2002}, but this lacks the diagrammatic notation and is more state-based rather than relation-based.

Model checking  is well known to sometimes produce a large number of states, which limits the scale of systems we can analyse.
This can be mitigated in places, \eg via bounded model checking~\cite{DBLP:conf/tacas/BiereCCZ99} and executing models at runtime~\cite{calinescu.ea_FormalMethodsAtRuntime:2010} (to only analyse the \emph{actual} path, not all possible paths).
Bigraphs are an active area of research and it is likely new analysis techniques will become available in future. For example, the theory already includes methods for computing equivalences (bisimulations) between agents (bigraphs)~\cite{DBLP:conf/concur/LeiferM00} that could be used to reduce the search space, but no implementation exists.
Likewise techniques from graph rewriting such as confluence checking~\cite{plump_HypergraphRewriting:1993} and inductive proofs~\cite{dyck_inductiveGTS:2015} would enable a much wider range of systems to be modelled.

\section{Conclusions} \label{sec:conclusions}

Bigraphs are a versatile modelling formalism, capable of describing a wide range of systems and their dynamics.
At their core is the juxtaposition of two relations: placement, described by a place graph (a forest), and connectivity, described by a hypergraph. The ability to connect entities in multiple ways not only models realistic scenarios, in that what you can do might depend both on \emph{where} you are, and who you might be \emph{connected} to, but leveraging both together can also overcome modelling challenges such as fixed arity constraints, ordering children, as well as  enabling   multi-perspective modelling. 

Bigraphical reactive systems (BRSs) allow models to evolve over time.
Dynamics are specified through \emph{user specified} reaction rules that rewrite (sub-)bigraphs by bigraphs. We have shown how our extensions of priority, conditional, instantaneous, probabilistic, and stochastic rewriting, coupled with
control-entities such as tagging, make BRS a powerful and expressive modelling framework.  

Model analysis can be performed through simulation and model checking by generating a transition system with bigraphs as states  and reaction rules as transitions. 

This tutorial has foregone much of the theoretical aspects of bigraphs in favour of practical modelling techniques;  key contributions are our Modelling Tips and examples expressed in BigraphER.

As Milner noted:

\begin{quote}
    ``The model [bigraphs] is only a proposal; it can only become a foundational model for
    ubiquitous computing if it survives serious experimental application. For
    the latter, it must be seen to yield language for programming and
    simulation, and equipped with appropriate mechanised tools for analysis,
    such as model checking.'' \cite{miler_BigraphLectureNotes:2008}
\end{quote}

With BigraphER, the programming and tooling support is now available:  it is time for the \emph{application}.
We hope this tutorial enables modellers to  learn the formalism quickly, benefit from bigraphs in their work, and  develop these ``serious'' experimental applications.

\section*{Acknowledgements}
This work is supported by the Engineering and Physical Sciences Research Council, under PETRAS SRF grants MAGIC and FARM (EP/S035362/1), S4: Science of Sensor Systems Software (EP/N007565/1) and an Amazon Research Award on Automated Reasoning. 

\appendix
\section{Modelling Tips}\label{tips} 
\begin{enumerate}
    \item 
 || and | allow us to build bigger bigraphs from smaller. Use || to model distinct bigraphs and | for merging bigraphs.

 \item Use 1 to indicate “no possibility of any children”  and id to indicate “zero or more children”.

 \item Use prefixes/suffixes in entity names (in textual format) and colours and shading (in diagrammatic format) to indicate relationships between states or stages of a process.  

 \item  Use a named, open link to indicate “this link potentially connects elsewhere”,
and a closed link to indicate “only these entities are connected”. The specific names used do not
matter.
 \item  Sites are abstractions over bigraphs, i.e. they are bigraph variables that can be instantiated with a bigraph.  Sites should be used when defining general rules that apply in many situations.

\item Careful consideration needs to be given to duplicating sites when the bigraphs being
duplicated contain links: when   a site is duplicated that contains links,  the links remain connected. For example, if we copy \biginline{A\{x\}} to obtain \biginline{A\{x\} | A\{x\}}, then both  \biginline{A} entities are connected in the result.

 \item Parameterised rules are syntactic sugar for a set of underlying rules, so use
them sparingly as each new rule increases the work needed for system analysis. In practice,
this affects how you describe entities. 
For example, it is  better to define \biginline{Camera.ID(1)}, 
which can be abstracted by a site whenever the identifier is unimportant,\eg \biginline{Camera.id},
rather than \biginline{Camera(1)}  which requires a family of rules \emph{every} time a rule uses a \biginline{Camera}.

 \item Be careful when assigning rule priorities. Although priorities stop a general
case applying when a specific should be applied, it also stops a general case being applied to
any other matches.

 \item Conditional rules allow us to restrict instantiation of sites: instead of allowing arbitrary bigraphs, conditions allow only those that do/do not match a pattern.

 \item Where possible, it is better to choose a conditional rule over rule priorities. This is because conditions indicate the intent of a single rule, while priorities  define relationships across the whole set of rewrite rules.     

 \item  For multi-perspective modelling, it is useful to add a top-level entity that
allows a region to be referred to by name, e.g. Physical, Social.

 \item  If it is difficult to model with places, add more link structure; if it is difficult
to model with links, add more place structure.

 \item If you want to apply a rule sequence a fixed number of times, add tags to entities to indicate whether or not the sequence has been applied, and modify the rules so they introduce and then remove tags. 

 \item There may be several possible ways to model a system feature, e.g. tagging,
conditional rules, parameterisation, etc. It is good practice to employ one consistent approach
throughout the model.

 \item Probabilistic, stochastic, and action bigraphs control how rules are applied,
e.g. how often, but they do not affect the rules themselves, which retain the same rewriting mechanism as before.

\end{enumerate}
    
\bibliographystyle{ACM-Reference-Format}

\end{document}